\documentclass[preprint,12pt]{elsarticle}
\usepackage{epsfig,amsmath,amssymb,slashed,bm}
\usepackage[utf8]{inputenc}
\usepackage{graphicx,subfigure}
\usepackage{hyperref}
\usepackage{xcolor}
\def\!{\mskip-\thinmuskip}

\newcommand{\mean}[1]{\langle #1 \rangle}

\newcommand{\tr}{{\rm tr}}  

\newcommand{\D}{{\rm d}}

\newcommand{\I}{ \mathrm{i}}
\newcommand{\h}[1]{\widehat{#1}}
\newcommand{\E}{{\rm e}}
\newcommand{\de}{\partial}
\newcommand{\subs}[1]{_{\textup{#1}}}

\newcommand{\krm}{{\rm k}}
\renewcommand{\vec}[1]{\ensuremath{\mathchoice
                     {\mbox{\boldmath$\displaystyle\mathbf{#1}$}}
                     {\mbox{\boldmath$\textstyle\mathbf{#1}$}}
                     {\mbox{\boldmath$\scriptstyle\mathbf{#1}$}}
                     {\mbox{\boldmath$\scriptscriptstyle\mathbf{#1}$}}}}

\renewcommand{\l}{\left}

\hypersetup{pdfinfo={Title={Spin polarization in B},Author={Matteo Buzzegoli}}}
\journal{Nuclear Physics A}
\begin{document}

\begin{frontmatter}
\title{Spin polarization induced by magnetic field and \texorpdfstring{\\}{ }
the relativistic Barnett effect} 

\author{M. Buzzegoli}
\ead{mbuzz@iastate.edu}
\affiliation{organization={Department of Physics and Astronomy, Iowa State University},
            addressline={2323 Osborn Drive}, 
            city={Ames},
            postcode={50011}, 
            state={Iowa},
            country={USA}}
\begin{abstract}
First, I study the analogy between the magnetization of a material and the spin
polarization of particles in a fluid. Using the relativistic version of the
Barnett effect, \textit{i.e.} the magnetization of a material induced by mechanical rotation,
the spin polarization induced by thermal vorticity is obtained within a purely
classical model, where spin is treated as an intrinsic magnetic moment and
rotation is included as a non-inertial effect. I argue that since spin
polarization induced by thermal vorticity can be obtained in a classical theory,
it can not be dominated by quantum anomalies.
\newline
Second, the spin polarization induced by magnetic field is obtained for a fluid
at local thermal equilibrium using statistical quantum field theory. The obtained
formula is valid beyond the weak field approximation and when contributions from
the non-homogeneity of the magnetic field are small. The exact form of spin polarization
is studied for a free Dirac field at global equilibrium, and, like magnetic susceptibility,
it oscillates according to the de Haas - van Alphen effect.
\newline
Finally, I briefly review how magnetic field contributes to the difference between
the spin polarization of $\Lambda$ and $\bar{\Lambda}$ observed in heavy-ion collisions.
\end{abstract}
%




\end{frontmatter}

\section{Introduction}
Non-central heavy-ion collisions produce a plasma with large magnetic
field \cite{Tuchin:2013ie} and angular momentum \cite{Becattini:2007sr}.
Both these two factors can be revealed in the measurements of $\Lambda$ spin
polarization \cite{STAR:2017ckg, Adam:2018ivw, Adam:2019srw, ALICE:2019aid, STAR:2020xbm, STAR:2021beb}. The impact of the vorticity to the spin polarization is predominant compared
to the magnetic field because the former survives for a longer time due to the angular
momentum conservation. For this reason, many studies have focused in the spin polarization
induced by vorticity and it is now a well establish phenomenon, see for
instance \cite{Becattini:2020ngo,Buzzegoli:2022kyj} for a review. It is curious how
the effect of vorticity is explained in analogy with magnetization as a relativistic
Barnett effect \cite{Barnett:1915,Barnett:1935}, while the spin polarization of a
relativistic particle induced by magnetic field itself received less attention.

Motivated by the spin physics in heavy-ion collisions, recent
studies \cite{Koide:2012kx,Singh:2022ltu,Bhadury:2022ulr} have included the spin
degrees of freedom in the magneto-hydrodynamic equations. The spin polarization
induced by magnetic field was derived in
Refs. \cite{Gao:2012ix,Becattini:2016gvu,Yi:2021ryh} and was analyzed 
in \cite{Guo:2019joy,Xu:2022hql}. Differently from vorticity, a magnetic field
produces opposite spin polarization for $\Lambda$ and $\bar\Lambda$. Other effects
that might produce a difference in spin polarization in heavy-ion collisions
were quantified in \cite{Ryu:2021lnx,Wu:2022mkr}. Assessing the
intensity of magnetic field for these measurements is also important for
estimating the magnitude of the chiral magnetic effect \cite{Muller:2018ibh}.

The previous derivations of the spin polarization induced by magnetic field relied
on an analogy with vorticity \cite{Becattini:2016gvu} or they were based on kinetic
theory \cite{Gao:2012ix,Yi:2021ryh}. The main purpose of this work is to derive
the spin polarization in a full quantum relativistic framework. Indeed the
Quark–Gluon Plasma is composed by strongly interacting quantum fields that can
not be described with weakly interacting quasi-particles \cite{Adams:2012th}.
Therefore the results of kinetic theory can not be extended inside the plasma as
it describes the fields as quasi-particles between the collisions. Despite this,
the system is a fluid and the notion of local thermal equilibrium admits a full
quantum relativistic description. In this work the spin polarization of a massive
spin $1/2$ fermion is derived using the Zubarev formalism for the non-equilibrium
statistical operator \cite{Zubarev:1966,Zubarev:1979,vanWeert1982,Zubarev:1989su,Becattini:2014yxa,Buzzegoli:2018wpy,Becattini:2019dxo,Buzzegoli:2020ycf} and the exact form of the Wigner function in magnetic
field \cite{Sheng:2017lfu,Gorbar:2017awz}.

This paper is structured as follows. In Sec. \ref{sec:Magn}, I review the magnetization
and the non-relativistic and relativistic version of the Barnett effect. I also show
that even classical theory, i.e. describing the spin as an intrinsic magnetic moment
and describing the effect of rotation as a non-inertial effect, predicts a spin
polarization along the rotation of the system. In Sec. \ref{sec:LTEB}, I derive the
local thermal equilibrium in the presence of vorticity and magnetic field and I obtain
a formula for spin polarization. In sec. \ref{sec:Wigner}, I write the scalar and
axial part of the Wigner function of a fermion in external magnetic field.
In sec.~\ref{sec:SpinPol}, I derive the spin polarization vector induced by magnetic field;
I compare the full result with a calculation in the non-relativistic limit; I study
this polarization at global equilibrium and I discuss the experimental data for
heavy-ion collisions.

\section{Magnetization and the relativistic Barnett effect}
\label{sec:Magn}
In this section I review the magnetization of a magnetic material induced by an external
magnetic field and by a mechanical rotation, which is the Barnett effect, and I derive
the relativistic version of the Barnett effect.

Using the classical theory, one can describe a magnetic material as composed by non-interacting
particles with magnetic moment $\vec\mu$. Denoting with $n$ the number density of magnetic moments and
with $T$ the temperature, the magnetization $M$ induced by a magnetic field $B$ is
\begin{equation}
\label{eq:ClassicalMagn}
    M = n \mu L\left(\frac{\mu B}{T}\right),
\end{equation}
where $L$ is the Langevin function
\begin{equation}
L(x) = \coth(x)-\frac{1}{x}
    \simeq \frac{x}{3} +\mathcal{O}\left( x^2\right).
\end{equation}
It is well known that the classical magnetization (\ref{eq:ClassicalMagn}) can be derived using classical statistical mechanics.
It is sufficient to describe the energy of the  magnetic elements in the magnetic field with the Hamiltonian
\begin{equation}
\label{eq:EnergyB}
    H =-\vec{\mu}\cdot\vec{B} =-\mu B \cos\theta.
\end{equation}
Consequently, the partition function of this system is
\begin{equation}
\begin{split}
Z =& \int_0^{2\pi}\D\phi\int_0^\pi\D\theta\, \sin\theta \E^{\mu B \cos\theta}
 = 2\pi\int_{-1}^1 \D y\, \E^{\mu B \beta y}
 = 4\pi \frac{\sinh(\mu B\beta)}{\mu B \beta},
\end{split}
\end{equation}
where $\beta$ is the inverse of temperature $T$. From the partition function one can easily
obtain the average magnetic moment in the direction of the magnetic field as
\begin{equation}
\begin{split}
\mean{\vec{\mu}\cdot\hat{\vec{B}}} =& \frac{1}{Z} \int_0^{2\pi}\D\phi\int_0^\pi\D\theta\, \sin\theta
    |\mu\cos\theta|\E^{\mu B \cos\theta}
    =\frac{1}{Z\beta}\frac{\de Z}{\de B}
    =\frac{1}{\beta}\frac{\de \log Z}{\de B},
\end{split}
\end{equation}
which results in the magnetization (\ref{eq:ClassicalMagn}).

Similarly, magnetic materials can be magnetized by rotating them, a phenomenon known as
Barnett effect~\cite{Barnett:1915,Barnett:1935}. Historically, the magnetization induced by
rotation was obtained balancing the torques acting on the magnetic moments under rotation
\cite{Barnett:1915,Barnett:1935}. Today, a modern description \cite{Matsuo2015} can be
given in terms of non-inertial effects caused by rotation. In order to obtain the result,
it is indeed sufficient to observe that in the presence of an angular velocity $\vec{\omega}$
the energy of the particle is shifted due to the coupling of angular momentum, denoted by
$\vec{J}$, with rotation \cite{deOliveira:1962apw,Hehl:1990nf}:
\begin{equation}
\label{eq:omegaJ}
H \to H - \vec{\omega}\cdot \vec{J}\, .
\end{equation}
Note that this is a purely classical effect.

Consider then a molecular magnet as a single particle with charge $q$ and mass $m$
revolving in a close orbit about a much more massive nucleus of charge $-q$. With $r$
the radius vector and $\omega$ the angular velocity, the magnetic moment of the molecular
system is given by $\mu = \frac{1}{2} q r \omega^2$, and its angular momentum is
\begin{equation}
\vec{J} = m r^2 \vec{\omega}\, .
\end{equation}
According to Eq. (\ref{eq:omegaJ}) the energy of the magnetic element is
\begin{equation}
\label{eq:EnergyOmega}
H = - \vec{J} \cdot \vec{\omega} = - J \omega \cos\theta
    = - \mu \frac{\omega}{\gamma}\cos\theta,
\end{equation}
where $\gamma$ denotes the gyromagnetic ratio $\gamma=\mu / J$. The comparison of
Eqs. (\ref{eq:EnergyB}) and (\ref{eq:EnergyOmega}) gives the classical Barnett effect,
which states that a mechanical rotation of the material induces the same magnetization
as an effective magnetic field with intensity
\begin{equation}
\label{eq:BEffBarnett}
    B_{\rm Eff} = \frac{\omega}{\gamma}.
\end{equation}

This classical description can be easily extended to the quantum world by describing
the magnetic elements as quantum particles with spin $S$ and using the non-relativistic
quantum statistical operator at thermal equilibrium
\begin{equation}
\label{eq:StatOperNonRel}
\h{\rho} = \frac{1}{Z}\exp\left[-\frac{\h{H}}{T}
    + \frac{\mu_0}{S T}\vec{B}\cdot \h{\vec{S}}
    + \frac{\vec{\omega}}{T} \cdot \left(\h{\vec{L}}+\h{\vec{S}}\right) \right],
\end{equation}
where the second term in the exponent is the Pauli interaction with an external magnetic field,
the last one is the angular momentum-angular velocity coupling and
$\h{\vec{\mu}}=\frac{\mu_0}{S}\hat{\vec{S}}$ is the magnetic moment of the particle with
$\mu_0=q/2m$. Consider the simple case where $\vec{B}$ is parallel to $\vec{\omega}$,
then the statistical operator can be diagonalized in the basis of the eigenvectors of the spin operator
component parallel to $\vec{B}$. The magnetization of the material is given in terms of the expectation 
value of $\h{\vec{\mu}}$, that is proportional to the spin in the direction of magnetic field:
\begin{equation}
\label{eq:SBExpe}
\begin{split}
\mean{\h{\vec{S}}\cdot \hat{\vec{B}}} = &
    \hat{\vec{B}} \frac{\sum_{\sigma=-S}^S \sigma\exp\left[\frac{\mu_0 B /S +\omega}{T}\sigma\right]}{\sum_{\sigma'=-S}^S \exp\left[\frac{\mu_0 B /S +\omega}{T}\sigma'\right]}
= \hat{\vec{B}} \frac{\frac{\de}{\de x}\sum_{\sigma=-S}^S \E^{x \sigma}}{\sum_{\sigma'=-S}^S \E^{x \sigma'}}
= S B_S(x),
\end{split}
\end{equation}
where I defined $x=(\omega + \mu_0 B/S)/T$, and $B_S$ is the Brillouin function:
\begin{equation}
\label{eq:BrillouinB}
\begin{split}
S B_S(x) =& \frac{2S+1}{2} \coth\left(\frac{2S+1}{2}x\right)-\frac{1}{2}\coth\left( \frac{x}{2}\right)
    \simeq \frac{S(S+1)}{3} x +\mathcal{O}\left( x^2\right).
\end{split}
\end{equation}
Setting the rotation to zero and considering a case where $x\ll 1$, the previous expressions
reproduce the Curie law for the magnetization:
\begin{equation}
\begin{split}
M =& n \mean{\h{\vec{\mu}}\cdot \hat{\vec{B}}}
    = \frac{n \mu_0}{S} \mean{\h{\vec{S}}\cdot \hat{\vec{B}}}
    \simeq n\, \mu_0 \frac{S+1}{3} x
    = n\, \mu_0^2 \frac{S+1}{3S}\frac{B}{T},
\end{split}
\end{equation}
which for $S=1/2$ is $M=n\beta \mu_0^2 B$. The Barnett effect is obtained turning on the rotation.
Indeed, the effect of rotation can be read by looking at the argument of the Brillouin function
in Eq. (\ref{eq:SBExpe}): the magnetization induced by rotation is the same as the one obtained
by an effective magnetic field (\ref{eq:BEffBarnett}) with a gyromagnetic ratio of $\gamma=\mu_0/S$.
The classical result (\ref{eq:ClassicalMagn}) corresponds to the limit $S\to\infty$ of
Eq. (\ref{eq:BrillouinB}), that reproduces the Langevin function
\begin{equation}
\lim_{S\to \infty} S B_S(x) = L(x).
\end{equation}
%

\subsection{The relativistic Barnett effect}
\label{sec:anomaly}
Since in the quantum approach the magnetization was obtained evaluating the expectation value
of the spin, it is clear that the phenomenon of magnetization is analogous to the one of spin
polarization. Despite the fact that the spin of a particle is an inherently quantum property,
the main reason that cause the magnetization is classical. Because of that, the spin
polarization induced by magnetic field and a parallel rotation can be obtained with the classical
theory described above and it is given by
\begin{equation}
\label{eq:SpinPolClassical}
\mean{\h{\vec{S}}\cdot \hat{\vec{B}}}= L(x),\quad
    x=\frac{\mu}{T}\left(B + \frac{\omega}{\gamma} \right).
\end{equation}
More accurately, for a quantum system in the non-relativistic limit, the spin
polarization is given by:
\begin{equation}
\mean{\h{\vec{S}}\cdot \hat{\vec{B}}}= S B_S(x),\quad
    x=\frac{\omega + \mu_0 B/S}{T} .
\end{equation}

To maintain the right covariant properties, the spin polarization for a relativistic particle
is given by a mean spin (four-)vector. Recently, the polarization of the spin of a particle in a
relativistic fluid has been studied intensively, see  for instance the reviews \cite{Becattini:2020ngo,Buzzegoli:2022kyj} and
reference therein. Around a point $x$ of the fluid, the spin vector of a particle with
momentum $p$ can be induced by the thermal vorticity of the fluid, which is defined as
\begin{equation}
\label{eq:ThermalVort}
\varpi_{\mu\nu} = -\frac{1}{2}\left(\de_\mu \beta_\nu - \de_\nu \beta_\mu\right),
\end{equation}
with $\beta^\mu=u^\mu/T$ and $u$ is the fluid velocity. For a spin $1/2$ particle the
resulting spin polarization is \cite{Becattini:2013fla,Becattini:2016gvu}
\begin{equation}
S^\mu_\varpi (x,p) = -\frac{1}{8m}(1-n\subs{F})\epsilon^{\mu\rho\sigma\tau}p_\tau\varpi_{\rho\sigma},
\end{equation}
where $n\subs{F}$ is the Fermi-Dirac thermal distribution function. Being an anti-symmetric
tensor, the thermal vorticity can be decomposed as
\begin{equation}
\varpi_{\mu\nu}= u_\nu \frac{a_\mu}{T} - u_\mu \frac{a_\nu}{T}
    + \epsilon_{\mu\nu\rho\sigma} \frac{\omega^\rho}{T} u^\sigma,
\end{equation}
where $a$ and $\omega$ are respectively the acceleration and the angular velocity of
the fluid \cite{Buzzegoli:2017cqy}. In a case without acceleration ($a=0$), the previous
expression reduces to:
\begin{equation}
\label{eq:LocalSpinVort}
S^\mu_\omega (x,p) = \frac{1}{4m}(1-n\subs{F})\beta\varepsilon_p (\omega^\mu-u^\mu \frac{\omega\cdot p}{\varepsilon_p}),
\end{equation}
where $\varepsilon_p=p\cdot u$.
As I will show in next Sections, the spin polarization induced by a magnetic field is
\begin{equation}
\label{eq:LocalSpinB}
S^\mu_B (x,p) = \frac{1}{4m}(1-n\subs{F})\beta (q B^\mu-u^\mu \frac{qB\cdot p}{\varepsilon_p}).
\end{equation}
The relativistic Barnett effect is derived comparing the two expressions.
The spin polarization induced by rotation is the same as the one induced
by a magnetic field with intensity
\begin{equation}
\label{eq:BEffBarnettRel}
    B^\mu_{\rm Eff} = \frac{\varepsilon_p}{q}\omega^\mu.
\end{equation}
This is the magnetic field required to produce a synchrotron frequency
equals to $\omega$.
In the non-relativistic limit this  effective magnetic field reduces to
\begin{equation}
B_{\rm Eff} = \frac{m}{q}\omega=\frac{1}{2\mu_0}\omega,
\end{equation}
which is the Barnett effect derived above in Eq. (\ref{eq:SBExpe}) for spin $S=1/2$.

\subsection{Spin polarization and gravitational anomaly}
The vorticity induced spin polarization is explained above as a relativistic Barnett effect.
It is possible to connect the spin polarization (or the magnetization) induced by a
magnetic field with the one induced by rotation thanks to the analogy of the Pauli
interaction with the energy shift due to rotation: compare the second and third term
in Eq. (\ref{eq:StatOperNonRel}). As the Pauli interaction is valid only in the
non-relativistic approximation, it is not expected that higher order corrections in
angular velocity can be obtained using the effective magnetic field (\ref{eq:BEffBarnettRel}).
In any case, as shown in \cite{Buzzegoli:2021jeh}, the ultimate reason why a spin
polarization is generated from rotation is the angular momentum-rotation coupling
in Eq. (\ref{eq:omegaJ}), which is a classical non-inertial effect. Indeed, it was
possible to obtain the pure classical result (\ref{eq:SpinPolClassical}) for the
spin polarization.

It was realized that for a massless field the thermal coefficient giving the linear
response of a spin current generated by rotation is tightly connected with the quantum
gravitational anomaly coefficient appearing in the divergence of the axial current
\cite{Landsteiner:2011cp}. Subsequently, it was proposed that the spin polarization
induced by vorticity might be caused by the gravitational anomaly \cite{Landsteiner:2011cp,Jensen:2012kj,Stone:2018zel,Prokhorov:2022udo}.
The connection with the quantum anomaly offers new insights in the interplay between quantum
field theory, transport and non-inertial effects, but, as argued above, I believe that the
cause of this phenomenon is the angular momentum and rotation coupling. If the quantum
gravitational anomaly were truly necessary, it could not have been possible to obtain a
spin polarization with a classical theory. Furthermore, a direct connection with the
gravitational anomaly is only possible for massless fields and breaks down for massive
fields, where the spin polarization has been actually measured.

\section{Local thermal equilibrium in magnetic field}
\label{sec:LTEB}
The main purpose of this work is to obtain the spin polarization of a free fermion in
a fluid at local thermal equilibrium in the presence of an external magnetic field.
The spin polarization of a fermion is obtained using \cite{Becattini:2020sww}
\begin{equation}
\label{eq:SpinPolFormula}
S^\mu(p)= \frac{1}{2}\frac{\int_{\Sigma} {\rm d}\Sigma\cdot p\; {\rm tr}_4\left[\gamma^\mu \gamma^5 W_+(x,p) \right] }
 {\int_{\Sigma} {\rm d}\Sigma\cdot p\; {\rm tr}_4 \left[W_+(x,p)\right]},
\end{equation}
where $W_+(x,p)$ is the particle branch of the Wigner function. In heavy-ion collision
the predictions for $\Lambda$ spin polarization are obtained by integrating the previous formula
over the freeze-out hypersurface. For a Dirac field in external electromagnetic field the
Wigner function is~\cite{Vasak:1987um}
\begin{equation}
\label{eq:WignerFunc}
\begin{split}
W(x,p)_{AB} =& \frac{1}{(2\pi)^4} \! \int \!{\rm d}^4 y\, {\rm e}^{-{\rm i} p \cdot y}\\
	&\times\langle \bar{\Psi}_B (x +y/2) U(A,x+y/2,x-y/2) \Psi_A (x-y/2) \rangle,
\end{split}
\end{equation}
where $U$ is the gauge link
\begin{equation}
    U(A,x_+,x_-)= \exp\left[-\I\int_{x_-}^{x_+}\D z^\mu A_\mu (z)\right],
\end{equation}
and the brackets denotes the thermal average with the statistical operator $\hat{\rho}$:
\begin{equation}
\langle \hat{X} \rangle = {\rm tr}\left(\hat{\rho}\,\hat{X}\right).    
\end{equation}
In what follows, I derive the statistical operator at local thermal equilibrium.
Then, I will use it to obtain the Wigner function (\ref{eq:WignerFunc}) and finally
I will use the Wigner function to obtain the spin polarization with Eq. (\ref{eq:SpinPolFormula}).

The current best model to describe the QGP created in heavy-ion collisions assumes
that the system reaches the local thermal equilibrium at a certain initial time.
A covariant description of a quantum system that reached the local thermal equilibrium is
given by the Zubarev method for the non-equilibrium statistical operator, see for instance
\cite{Hayata:2015lga,Becattini:2019dxo} for a review. In this approach the statistical
operator is obtained by maximizing the total entropy at fixed energy-momentum 
and charge density in the initial time hypersurface. The local equilibrium statistical
operator at later times is obtained by evolving the system and by neglecting the dissipative
effects, and it is given by:
\begin{equation}
\label{eq:RhoLTE}
\h{\rho}\subs{LTE} = \frac{1}{Z} \exp \left[ \int_\Sigma \!\!\D\Sigma_\mu(y) \left(
    \h{T}^{\mu\nu}(y) \beta_\nu(y) - \zeta(y) \h{j}^\mu(y) \right)\right],
\end{equation}
where $\Sigma$ is the hypersurface at a given time, $\h{T}_{\mu\nu}$ is the symmetric
stress-energy tensor (SET), $\h{j}^\mu$ is the conserved electric current, $\beta^\mu=u^\mu/T$ is
the four-temperature vector and $\zeta$ is the ratio of chemical potential and temperature.

Furthermore, the magnetic field in heavy-ion collisions can be treated as an external
field, as its dynamics approximately decouple from that of the hot nuclear
medium \cite{Stewart:2017zsu}. In the presence of an external electromagnetic field the
statistical operator (\ref{eq:RhoLTE}) remains the same, but the SET is no longer
conserved because of the Lorentz force:
\begin{equation}
\label{eq:DivergenceSET}
\de^\mu \h{T}_{\mu\nu} = \h{j}^\lambda F_{\nu\lambda},\quad
\de^\mu \h{j}_\mu = 0.
\end{equation}
This approach was used to obtain the relativistic magneto-hydrodynamic equations
\cite{Huang:2011dc,Hongo:2020qpv} and was recently studied in \cite{Buzzegoli:2020ycf}
with the mutual presence of thermal vorticity and electromagnetic field.

Notice that since the statistical operator (\ref{eq:RhoLTE}) has the same form with
and without an external electromagnetic field, it is important to include the external
gauge field inside the operators and solve the equations of motion for the Dirac field
in external electromagnetic field. For a free Dirac field interacting with an external
gauge field $A^\mu$, the symmetric (and gauge invariant) stress-energy tensor is
\begin{equation}
\label{eq:SETwithA}
\begin{split}
\h T^{\mu\nu}=&\frac{\I}{4}\left[\bar{\psi}\gamma^\mu\overrightarrow{\de}^\nu\psi
    -\bar{\psi}\gamma^\mu\overleftarrow{\de}^\nu\psi
    +\bar{\psi}\gamma^\nu\overrightarrow{\de}^\mu\psi
    -\bar{\psi}\gamma^\nu\overleftarrow{\de}^\mu\psi\right]\\
&-\frac{1}{2} \left(\h j^\mu A^\nu+\h j^\nu A^\mu\right).
\end{split}
\end{equation}
Using the equations of motion, one can check that this SET satisfies the
condition in Eq. (\ref{eq:DivergenceSET}).

The system is at global thermal equilibrium when the statistical operator (\ref{eq:RhoLTE})
becomes stationary. This occurs when the operator does not depend on the hypersurface,
or equivalently when the divergence of the integrand is vanishing \cite{Becattini:2012tc}. 
Taking advantage of Eq. (\ref{eq:DivergenceSET}), one finds that the global equilibrium
is realized when the thermodynamics fields solve the following equations:
\begin{equation}
\label{eq:GlobalEq}
    \de_\mu\beta_\nu + \de_\mu\beta_\nu = 0,\quad
    \de^\mu\zeta = F^{\nu\mu} \beta_\nu.
\end{equation}
The first one is a Killing equation and its solution in flat space is
\begin{equation}
\beta^\mu(y) = b^\mu + \varpi^{\mu\nu} y_\nu,
\end{equation}
where $b$ is a constant time-like vector and $\varpi$ is a constant thermal
vorticity, defined in Eq. (\ref{eq:ThermalVort}). For a relativistic system the
general global thermal equilibrium is realized with a non-homogeneous temperature and
with a constant non-vanishing vorticity. A global equilibrium is possible even in the
mutual presence of vorticity and electromagnetic field if the electromagnetic strength tensor
satisfies
\begin{equation}
\label{eq:GlobalEqF}
\beta_\sigma(y) \de^\sigma F^{\mu\nu}(y) = \varpi^\mu_{\;\sigma} F^{\sigma\nu}(y)
    - \varpi^\nu_{\;\sigma} F^{\sigma\mu}(y) .
\end{equation}
In this case, a general solution for the second Eq. in (\ref{eq:GlobalEq})
is given by \cite{Buzzegoli:2020ycf}
\begin{equation}
\label{eq:GeneralZetaGlobal}
\zeta(y) = \zeta_0 - \beta_\sigma(y) A^\sigma(y) - \Phi(y),
\end{equation}
where $\Phi$ is necessary to maintain the gauge invariance. For instance
a global equilibrium configuration is realized in a constant electromagnetic field
such that $F^{\mu\nu}=k\varpi^{\mu\nu}$, with $k$ a constant. In this configuration
the solution (\ref{eq:GeneralZetaGlobal}) becomes
\begin{equation}
\zeta(y) = \zeta_0 - \beta_\sigma(y) F^{\lambda\sigma}y_\lambda
    + \frac{1}{2}\varpi_{\rho\sigma} y^\rho F^{\lambda\sigma} y_\lambda .
\end{equation}
In Section \ref{sec:GlobalEqPol} the spin polarization induced by magnetic field is
studied for a system at global equilibrium. Since the focus of this work is the
effect of magnetic field,
the vorticity is set to zero and the global equilibrium conditions result in a constant
homogeneous electromagnetic field, as required by Eq. (\ref{eq:GlobalEqF}). Considering
the case of a vanishing electric field the global thermal equilibrium statistical operator
in the presence of an external magnetic field assumes the form \cite{Buzzegoli:2020ycf}
\begin{equation}
\label{eq:GlobalEqB}
\h{\rho}\subs{GTE,B} = \frac{1}{Z} \exp\left\{-\beta\cdot \h{P} + \zeta \h{Q}  \right\}
\end{equation}
where $\beta$ is a constant time-like vector, $\zeta$ is also constant, $\h{Q}$ is
the total electric charge and $\h{P}$ is the four-momentum of the system. As in the
local equilibrium case, the global statistical operator has the same form with or without
an external magnetic field. The effect of magnetic field must be included directly
in the field operators of the Dirac field.

Going back to the out of equilibrium case, the Wigner function at local thermal
equilibrium is given by
\begin{equation}
\label{eq:WignerLE}
\begin{split}
W(x,p)\subs{LTE}=& \frac{1}{Z} {\rm tr} \left[ \exp\left\{
-\int_\Sigma\!\! {\rm d}\Sigma_\mu(y)\left(\widehat{T}^{\mu\nu}(y)\beta_\nu(y)
 -\zeta(y)\,\widehat{j}^\mu(y)\right)\right\} \widehat{W}(x,p) \right].
\end{split}
\end{equation}
For a fluid in the hydrodynamic regime, such as the QGP, the local equilibrium Wigner function
is well approximated by expanding the thermodynamic fields in the exponent of (\ref{eq:WignerLE}).
Since the Wigner function is to be evaluated at the point $x$, the thermodynamic fields and the
gauge field are expanded around $x$:
\begin{equation}
\label{eq:TaylorBeta}
\begin{split}
\beta_\nu(y) \simeq& \beta_\nu(x) + \partial_\lambda \beta_\nu(x) (y-x)^\lambda,\\
\zeta(y) \simeq& \zeta_\nu(x) + \partial_\lambda \zeta(x) (y-x)^\lambda\, ,\\
A_\nu(y) \simeq& A_\nu(x) + \partial_\lambda A_\nu(x) (y-x)^\lambda.
\end{split}
\end{equation}
Notice that Eq. (\ref{eq:WignerLE}) depends on the gauge field through the SET (\ref{eq:SETwithA}).
For the sake of simplicity, the external electromagnetic field is considered as
composed only of a magnetic component, that is
\begin{equation}
\label{eq:ExternalB}
\de_\mu A_\nu - \de_\nu A_\mu = -\epsilon_{\mu\nu\rho\sigma}B^\rho u^\sigma.
\end{equation}
With this expansion the statistical operator is approximated by
\begin{equation}
\label{eq:ApproxHydro}
\begin{split}
\widehat{\rho}_{LTE} \simeq \frac{1}{Z} \exp& \left[ - \beta_\nu(x) \widehat{P}^\nu +\zeta(x)\widehat{Q}
	+ \frac{1}{2} \varpi_{\mu\nu}(x) \widehat{J}^{\mu\nu}_x \right.\\
    &\left.+\de_\lambda \zeta(x)\! \int\! {\rm d} \Sigma_\mu (y-x)^\lambda \widehat{j}^{\mu}(y)
	-\frac{1}{2} \xi_{\mu\nu}(x) \widehat{Q}^{\mu\nu}_x+\cdots\right],
\end{split}
\end{equation}
where
\begin{equation}
\label{eq:OperatorJ}
\widehat{J}^{\mu\nu}_x =\! \int \!{\rm d} \Sigma_\lambda \left[ (y-x)^\mu \widehat{T}^{\lambda\nu}(y) -
  (y-x)^\nu \widehat{T}^{\lambda\mu}(y)\right]
\end{equation}
is the generator of boosts and rotations (the conserved angular momentum operator) and
\begin{equation}
\label{eq:OperatorQ}
\widehat{Q}^{\mu\nu}_x =\! \int \!{\rm d} \Sigma_\lambda \left[ (y-x)^\mu \widehat{T}^{\lambda\nu}(y) + 
  (y-x)^\nu \widehat{T}^{\lambda\mu}(y)\right]
\end{equation}
is a non-conserved symmetric quadrupole like operator. I also introduced the thermal shear
\begin{equation}
\xi_{\mu\nu}=\frac{1}{2}\left(\partial_\mu\beta_\nu + \partial_\nu\beta_\mu \right).
\end{equation}
The presence of thermal vorticity, thermal shear and gradients of the chemical
potential induces a spin polarization
\begin{equation}
\label{eq:SpinPolFirstOrder}
\begin{split}
S^\mu(p) =& -\epsilon^{\mu\rho\sigma\tau} p_\tau \frac{\int_{\Sigma} {\rm d} \Sigma \cdot p \; n\subs{F} (1 -n\subs{F}) \mathcal{S}_{\rho\sigma} }
  { 8m \int_{\Sigma} {\rm d} \Sigma \cdot p \; n\subs{F}},\\
\mathcal{S}_{\rho\sigma} \equiv & \varpi_{\rho\sigma}
    + 2\, \hat t_\rho \frac{p^\lambda}{\varepsilon_p} \xi_{\lambda\sigma}
    -\frac{\hat{t}_\rho\partial_{\sigma}\zeta}{2\varepsilon_p},
\end{split}
\end{equation}
where $\hat{t}$ is the time direction in the laboratory frame and
$n\subs{F}=n\subs{F}(\beta(x)\cdot p-\zeta(x))$ is the Fermi-Dirac distribution function.
The first term is the known vorticity induced polarization \cite{Becattini:2013fla},
the second is the recently derived shear induced
polarization \cite{Becattini:2021suc,Becattini:2021iol,Liu:2021uhn,Fu:2021pok,Yi:2021ryh},
and the last one is the contribution from the gradient of the chemical potential that has been
studied in \cite{Liu:2020dxg,Fu:2022myl,Ivanov:2022geb}.
In heavy-ion collisions these effects give a larger contribution to spin polarization compared
to the contribution of the magnetic field. Indeed the magnetic induced polarization is of
opposite sign for $\Lambda$s and $\bar\Lambda$s but the measured polarization is roughly
the same.

Notice that it turned out that the magnetic field in~(\ref{eq:ExternalB}) does not explicitly
appear in the statistical operator (\ref{eq:ApproxHydro}). As in the case of global equilibrium
(\ref{eq:GlobalEqB}) the effect of the external magnetic field is included directly in the Dirac
field operators. If second order terms in derivatives in the expansion (\ref{eq:TaylorBeta}) of
the gauge field were included, then the statistical operator approximation (\ref{eq:ApproxHydro})
would have included a term proportional to $\de_\mu B_\nu(x)$. With the current data it might be
possible to observe the effects of magnetic field, but the effects of its gradients are smaller.
I therefore left the inclusion of such effects for future analysis. The leading order effect
of magnetic field in the Wigner function is obtained using the statistical operator
\begin{equation}
\label{eq:ApproxHydroB}
\begin{split}
\widehat{\rho}_{LTE} \simeq \frac{1}{Z} \exp& \left[ - \beta_\nu(x) \widehat{P}^\nu
    +\zeta(x)\widehat{Q}\right].
\end{split}
\end{equation}
One could also search for an interplay between magnetic field and the thermal vorticity and
shear by studying the operators (\ref{eq:OperatorJ}) and (\ref{eq:OperatorQ}) in the presence
of a magnetic field, but as they are higher order terms I leave those for future works. 
The spin polarization induced by magnetic field is then obtained using Eq. (\ref{eq:SpinPolFormula})
with the Wigner function (\ref{eq:WignerFunc}) evaluated with the statistical
operator (\ref{eq:ApproxHydroB}). Because the result is obtained expanding the fields
around a fixed point $x$, the resulting Wigner function is the same as a Wigner function
obtained at the global equilibrium (\ref{eq:GlobalEqB}) with a constant four-temperature and
magnetic field with values $\beta^\mu(x)$ and $B^\mu(x)$.

Note that there is an important difference between the Quantum Kinetic Theory (QKT)
and the Zubarev method discussed here. Whereas the Wigner function (\ref{eq:WignerLE}) is evaluated
from the statistical operator derived by maximizing the entropy at an initial time, QKT can only be carried out by providing a distribution function. In a underlying quantum field theory the correct
distribution function is not derived with a kinetic theory but the distribution function is provided
as an educated guess. Instead, the Wigner function derived here is a consequence of local thermal
equilibrium and it is valid in a full quantum relativistic regime. The distribution function might
be difficult to guess as for the case of equilibrium with thermal vorticity \cite{Palermo:2021hlf}.

\section{Wigner function in magnetic field}
\label{sec:Wigner}
In the previous Section it was found that spin polarization induced by magnetic field
is obtained with
\begin{equation}
\label{eq:SpinPolFormulaB}
S^\mu_B(p)= \frac{1}{2}\frac{\int_{\Sigma} {\rm d}\Sigma\cdot p\; \mathcal{A}_+(x,p) }
 {\int_{\Sigma} {\rm d}\Sigma\cdot p\; \mathcal{F}_+(x,p)},
\end{equation}
where
\begin{equation}
\begin{split}
\mathcal{F}_+(x,p)=&\tr\left[W_+(x,p) \right],\\
\mathcal{A}_+^\mu(x,p)=&\tr\left[\gamma^\mu\gamma^5 W_+(x,p) \right],
\end{split}
\end{equation}
are respectively the scalar part and the axial part of the particle branch of the Wigner
function (\ref{eq:WignerFunc}) obtained with the statistical operator
(\ref{eq:ApproxHydroB}) in external magnetic field.

The Wigner function in external magnetic field is obtained by first solving the Dirac
equation in magnetic field and then using that solution to evaluate the Wigner function.
The solutions of the Dirac equation in magnetic field are well known, see for
instance \cite{book:SokolovAndTernov,Buzzegoli:2020ycf}. The equilibrium Wigner function in external
magnetic field was already obtained using statistical quantum field theory~\cite{Hakim:1977},
but the spin and magnetic moment part were neglected because in astrophysical situations they
were negligible. Recently, the full Wigner function in external magnetic field was obtained
with exact methods in~\cite{Sheng:2017lfu,Gorbar:2017awz}.

Spin polarization only requires the scalar and axial part of the Wigner function.
In what follows I omit the $x$ dependence which is contained inside the temperature,
the chemical potential and the magnetic field. The scalar and axial parts of Wigner
function (\ref{eq:WignerFunc}) are given by \cite{Sheng:2017lfu}
\begin{align}
\label{eq:ExactWignerF}
\mathcal{F}_+(p)=&\sum_{n=0}^\infty \frac{4m}{(2\pi)^3}
    \theta(\varepsilon_p)\delta\left(\varepsilon_p^2 - {E_{p_z}^{(n)}}^2\right)\E^{-\xi/2}\\
    &\times n\subs{F}(\beta\cdot p-\zeta)
    (-1)^n \left[ L_n(\xi) -L_{n-1}(\xi) \right],\nonumber\\
\label{eq:ExactWignerA}
\mathcal{A}_+^\mu(p)  =& \sum_{n=0}^\infty \frac{4 a^\mu}{(2\pi)^3}
    \theta(\varepsilon_p)\delta\left(\varepsilon_p^2 - {E_{p_z}^{(n)}}^2\right) \E^{-\xi/2}\\
    &\times n\subs{F}(\beta\cdot p-\zeta)
    (-1)^n \left[ L_n(\xi) + L_{n-1}(\xi) \right],\nonumber
\end{align}
where $n$ denotes the Landau level, $L_n(z)$ is a Laguerre polynomial,
\begin{equation}
\varepsilon_p=p\cdot u,\quad
\xi=\frac{2 p_T^2}{|qB|},\quad
a^\mu = \varepsilon_p\frac{qB^\mu}{|qB|}-\frac{qB\cdot p}{|qB|}u^\mu,
\end{equation}
with $p_z$ and $p_T$ denoting respectively the parallel and transverse
(respect to the magnetic field) components of  momentum, and the energy spectrum is
\begin{equation}
    E_{p_z}^{(n)}=\sqrt{2n|qB| + p_z^2 + m^2}.
\end{equation}
Notice that since the Wigner function (\ref{eq:WignerFunc}) is gauge invariant,
the solutions written above are gauge independent. With the Wigner function in
Eqs. (\ref{eq:ExactWignerF}) and (\ref{eq:ExactWignerA}) one can compute the
spin polarization at every order
of magnetic field. However, it is more convenient to obtain the leading orders
as a power series in magnetic field. I will consider two limits. The first, that
is more relevant for application, is the weak field approximation when the magnetic
field is weaker than the momentum of the particle and than the temperature
\begin{equation}
    |q B| \ll p^2, T^2 .
\end{equation}
The other one is the opposite case where the magnetic field is very strong.

Using the methods developed in \cite{Gorbar:2017awz}, one can obtain the Wigner function
as a series in magnetic field in the weak field limit. After expanding the Wigner function
in powers of the magnetic field and summing over the Landau levels (see \ref{sec:AppWeak}),
its scalar and axial part is approximated by
\begin{align}
\label{eq:WeakWignerF}
\mathcal{F}_+(p)=&
    \frac{4m}{(2\pi)^3} n\subs{F}(\beta\cdot p-\zeta)
        \theta(\varepsilon_p)\delta\left(p^2 -m^2\right)\\
    & + \mathcal{O}\left(|qB|^2\right), \nonumber\\
\label{eq:WeakWignerA}
\mathcal{A}_+^\mu(p) =& \frac{4|qB|a^\mu}{(2\pi)^3}n\subs{F}(\beta\cdot p-\zeta)
    \theta(\varepsilon_p)\delta'(p^2-m^2)\\
    &+ \mathcal{O}\left(|qB|^2\right). \nonumber
\end{align}
To the best of my knowledge, this is the first
time this result has been derived from quantum field theory without relying on a
kinetic theory. It is in agreement with was previously found using quantum kinetic
theory in the Wigner formalism for
massless \cite{Gao:2012ix,Chen:2012ca,Hidaka:2017auj,Dong:2020zci,Yi:2021ryh} and
massive~\cite{Fang:2016vpj,Gao:2019znl,Weickgenannt:2019dks,Hattori:2019ahi,Liu:2020ymh,Dayi:2020uwx,Sheng:2020oqs,Gao:2020pfu}
fermions. Indeed it was found \cite{Gao:2019znl,Weickgenannt:2019dks}
\begin{equation}
\label{eq:AxialWignerKT}
\begin{split}
\mathcal{A}^\mu\subs{+,KT} =& \frac{q}{2}
    \frac{4\epsilon^{\mu\nu\rho\sigma} F_{\rho\sigma} p_\nu}{(2\pi)^3}
    n\subs{F}(\beta\cdot p-\zeta)\theta(\varepsilon_p)\delta'(p^2-m^2),
\end{split}
\end{equation}
which reduces to (\ref{eq:WeakWignerA}) using
$F_{\mu\nu}= -\epsilon_{\mu\nu\rho\sigma}B^\rho u^\sigma$.
It is straightforward to check that the integral over momentum of Eq. (\ref{eq:WeakWignerA})
results in the chiral separation effect:
\begin{equation}
j\subs{A\,+}^z = \int \D^4 p\, \mathcal{A}_+^z(p)
    =\frac{qB}{2\pi^2}\int_0^\infty \D p\, n\subs{F}(\beta E_p-\zeta),
\end{equation}
where $E_p =\sqrt{m^2 + \vec{p}^2}$. 
This is the same result one obtains using the full solution (\ref{eq:ExactWignerA}),
see for instance \cite{Sheng:2017lfu,Gorbar:2017awz,Buzzegoli:2020ycf}.

Instead, in the strong field limit $|q B| \gg p^2,\, T^2$, only the firsts Landau levels are
populated. Then, the Wigner function can be approximated by considering only the first
terms in the series (\ref{eq:ExactWignerF}) and (\ref{eq:ExactWignerA}). For instance,
considering only the Lowest Landau Level (LLL) $n=0$, it reads
\begin{align}
\label{eq:StrongWignerF}
\mathcal{F}_+(p)\subs{LLL}=& \frac{4m}{(2\pi)^3} n\subs{F}(\beta E_{p_z}^{(0)}-\zeta)\E^{-\xi/2}
    \frac{\delta(\varepsilon_p - E_{p_z}^{(0)})}{2E_{p_z}^{(0)}},\\
\label{eq:StrongWignerA}
\mathcal{A}_+^\mu(p)\subs{LLL}=& \frac{4 a^\mu}{(2\pi)^3}n\subs{F}(\beta E_{p_z}^{(0)}-\zeta)
    \E^{-\xi/2} \frac{\delta(\varepsilon_p - E_{p_z}^{(0)}) }{2E_{p_z}^{(0)}}.
\end{align}
%

\section{Spin polarization}
\label{sec:SpinPol}
The full spin polarization induced by magnetic field is obtained by plugging
the Eqs. (\ref{eq:ExactWignerF}) and (\ref{eq:ExactWignerA}) into (\ref{eq:SpinPolFormulaB})
and it is studied below at global equilibrium.

For practical applications and to understand what are the effects of the magnetic field,
it is more convenient to use the weak field approximation. Using
the Eqs. (\ref{eq:WeakWignerF}) and (\ref{eq:WeakWignerA}), the spin polarization
(\ref{eq:SpinPolFormulaB}) becomes
%
\begin{equation}
\label{eq:SpinB}
\begin{split}
S^\mu_B(p) =& \frac{\int_{\Sigma} \D \Sigma \cdot p \; \left(\varepsilon_p qB^\mu -u^\mu (qB\cdot p) \right) n\subs{F}(\beta\cdot p-\zeta)\theta(\varepsilon_p)\delta'(p^2-m^2)
 }{2m\int_{\Sigma} \D \Sigma \cdot p \; n\subs{F}(\beta\cdot p-\zeta)\theta(\varepsilon_p)\delta(p^2-m^2)}\\
 &+ \mathcal{O}\left(|qB|^2 \right).
\end{split}
\end{equation}
%
Notice that the spin vector is orthogonal to the momentum $p$; this is a property
inherited from the Pauli-Lubanski vector \cite{Becattini:2020sww}.

To remove the Dirac deltas and obtain on-shell momenta, I integrate the integrand of the
numerator and of the numerator of Eq. (\ref{eq:SpinB}) in $\varepsilon_p$. The denominator gives
\begin{equation*}
\int \!\D \varepsilon_p\, \delta(p^2-m^2)\theta(\varepsilon_p)n\subs{F}(\beta\cdot p -\zeta)
=\frac{n\subs{F}(\beta E_p-\zeta)}{2 E_p}.
\end{equation*}
In shorthand notation, the numerator becomes
\begin{equation*}
\begin{split}
N^\mu=&\int\D\varepsilon_p\, |qB|a^\mu  n\subs{F} \theta \delta'(p^2-m^2)
=-\int\D\varepsilon_p\, \delta(p^2-m^2) \frac{\D}{\D\varepsilon_p}
    \left[\frac{|qB|a^\mu  n\subs{F} \theta}{2\varepsilon_p}\right]\\
=&-\int\D\varepsilon_p\, \frac{\theta\delta(\varepsilon_p-E_p)}{2\varepsilon_p}
    \frac{\D}{\D\varepsilon_p}
    \left[\frac{|qB|a^\mu  n\subs{F} \theta}{2\varepsilon_p}\right].
\end{split}
\end{equation*}
Using
\begin{equation*}
\begin{split}
\frac{\D}{\D\varepsilon_p} \left( \frac{|qB| a^\mu}{\varepsilon_p}\right)
    =& \frac{qB\cdot p}{\varepsilon_p^2} u^\mu,\\
\frac{\D}{\D\varepsilon_p}n\subs{F}=&-\beta n\subs{F}(1-n\subs{F}),
\end{split}
\end{equation*}
one obtains
\begin{equation*}
N^\mu = \beta n\subs{F}(1-n\subs{F}) \frac{|qB| a^\mu}{4 E_p^2}
    - n\subs{F} \frac{qB\cdot p}{4 \varepsilon_p^3} u^\mu,
\end{equation*}
where the momentum is on-shell: $\varepsilon_p=E_p$.
Notice that the first term is orthogonal to the momentum $p$:
\begin{equation*}
    p_\mu a^\mu = \varepsilon_p\frac{qB\cdot p}{|qB|}-\frac{qB\cdot p}{|qB|}p\cdot u = 0,
\end{equation*}
whereas the second is not. But, as mentioned above, the spin vector must satisfy
$p_\mu S^\mu(p) =0$. Therefore, the second term must be vanishing when integrated
over the hypersurface. The final result is
\begin{equation}
\label{eq:PolWignerB}
S^\mu_B(p) = \frac{\int_{\Sigma}\! \D \Sigma \cdot p\, \beta(x) n\subs{F} (1 -n\subs{F})
 (qB^\mu(x) -u^\mu \frac{(qB(x)\cdot p)}{\varepsilon_p} )  }{4m\int_{\Sigma} \D \Sigma \cdot p \; n\subs{F}}
\end{equation}
where the momentum is on-shell and I denoted
\begin{equation}
    n\subs{F}=n\subs{F}\left(\beta(x)\cdot p-\zeta(x)\right)
\end{equation}
with $n\subs{F}(z)=(\E^{z}+1)^{-1}$. The same procedure described above can be
applied for the anti-particle branch of the Wigner function, leading to the result
\begin{equation}
\label{eq:PolWignerBant}
\bar{S}^\mu_B(p) = -\frac{\int_{\bar\Sigma}\! \D \Sigma \cdot p\, \beta \bar{n}\subs{F} (1 -\bar{n}\subs{F})
 (qB^\mu-u^\mu \frac{(qB\cdot p)}{\varepsilon_p})}{4m\int_{\bar\Sigma} \D \Sigma \cdot p \; \bar{n}\subs{F}},
\end{equation}
where $\bar{n}\subs{F}=n\subs{F}\left(\beta(x)\cdot p+\zeta(x)\right)$ and $\bar\Sigma$
in heavy-ion collisions denotes the freeze-out hypersurface of the anti-particle. The
main difference with the particle spin polarization is of course the overall change in
sign. This is in agreement with the chiral kinetic theory results of \cite{Gao:2012ix}
(in the massless limit) and with \cite{Yi:2021ryh} for a massive fermion (although
the term proportional to $u^\mu$ was omitted).

In experiments the spin polarization is measured in the rest frame of the particle
$\vec{S}^*$. This quantity is obtained from Eq. (\ref{eq:PolWignerB}) with the back boost:
\begin{equation}
\vec{S}^* = \vec{S}-\frac{\vec{p}\cdot\vec{S}}{p_0(p_0+m)}\vec{p}.
\end{equation}
Performing this operation, one can see that the part of Eq. (\ref{eq:PolWignerB})
proportional to $u$ does not contribute to $\vec{S}^*$.

It is now interesting to compare the result (\ref{eq:PolWignerB}) with the
spin polarization induced by rotation. From Eq. (\ref{eq:SpinPolFirstOrder}) keeping only the
rotational part of the thermal vorticity
$\varpi_{\rho\sigma}=\epsilon_{\rho\sigma\kappa\lambda}\beta\omega^\kappa u^\lambda$,
the spin polarization induced by rotation is
\begin{equation}
\label{eq:PolWignerOmega}
S^\mu_\omega(p) = \frac{\int_{\Sigma}\! \D \Sigma \cdot p\, \beta n\subs{F} (1 -n\subs{F})
 (\varepsilon_p \omega^\mu -u^\mu (\omega\cdot p) )  }{4m\int_{\Sigma} \D \Sigma \cdot p \; n\subs{F}}.
\end{equation}
Comparing the Eq. (\ref{eq:PolWignerB}) with the Eq. (\ref{eq:PolWignerOmega}),
one realizes that the spin polarization induced by rotation can be explained using
the relativistic Barnett effect with the effective magnetic field (\ref{eq:BEffBarnettRel})
as discussed in Sec. \ref{sec:anomaly}.

\subsection{The non-relativistic case}
In the non-relativistic limit, using $\frac{q}{2\varepsilon_p}\simeq \mu_0$,
the expression (\ref{eq:PolWignerB}) reduces to the known result \cite{Becattini:2016gvu}
\begin{equation}
\label{eq:PolNonRelSubstitution}
S^\mu\subs{B,NR}= \frac{\int_{\Sigma} \D \Sigma \cdot p \; \beta n\subs{F} (1 -n\subs{F}) 
 \mu_0\left(B^\mu\varepsilon_p -u^\mu (B\cdot p) \right)  }{2m\int_{\Sigma} \D \Sigma \cdot p \; n\subs{F}},
\end{equation}
where the spin polarization is given in terms of the magnetic moment of the particle.
The non-relativistic limit can be computed using linear response theory starting
from the statistical operator:
\begin{equation}
\label{eq:RhoB}
\h{\rho}\subs{B,NR} = \frac{1}{Z} \exp\left\{-\frac{\h{H}}{T} + \frac{\vec{\h{\mu}}}{T}\cdot \vec{B} \right\},
\end{equation}
where $\mu$ is the magnetic moment $\vec{\h{\mu}}=2\mu_0\vec{\h{S}}$.
This is essentially the same calculation of spin polarization induced by thermal vorticity
leading to Eq. (\ref{eq:SpinPolFirstOrder}). Then, the result (\ref{eq:PolNonRelSubstitution})
is simply obtained replacing\footnote{There is a minus sign between $\varpi$ and $F$, because rotation field and magnetic field are defined with a difference in sign.}
$\varpi_{\mu\nu}\to -2\beta\mu_0 F_{\mu\nu}$.
This spin polarization is the non-relativistic limit $\frac{q}{2\varepsilon_p}\simeq \mu_0$
of the polarization (\ref{eq:PolWignerB}).

Here I report the explicit calculation to check if the replacement actually gives the correct result.
The effect of the magnetic field on the Wigner function is obtained by applying the linear
response theory to the second term in the exponent of (\ref{eq:RhoB}):
\begin{equation*}
\begin{split}
\Delta W_+(x,p) =&  2\mu_0 B(x) \int\D^3 y\int_0^1\D z \,
        \mean{\h{W}_+(x,p) \h{\mathcal{S}}^z(y+\I z \beta)}_{c,\beta} ,
\end{split}
\end{equation*}
where the bracket $\mean{\cdots}_{c,\beta}$ denotes a connected thermal average with the
statistical operator (\ref{eq:RhoB}) with $\vec{B}=0$,
and the spin operator is obtained from the canonical spin tensor:
\begin{equation*}
\h{\mathcal{S}}^z(y) = \frac{\I}{2} \bar\Psi(y) \gamma^1 \gamma^2 \Psi(y).
\end{equation*}
Expanding the Dirac field in terms of the free creation-annihilation operators and using
\begin{equation*}
\begin{split}
\mean{\h{a}^\dagger_\tau(k) \h{a}_{\sigma}(q)}_{\beta}=&
    \delta_{\tau\sigma}2\varepsilon_q \delta^3(\vec{k}-\vec{q})n\subs{F}(\beta\cdot k),\\
\mean{\h{a}_{\tau'}(k')\h{a}^\dagger_{\sigma'}(q')}_{\beta}=&
    \delta_{\tau'\sigma'}2\varepsilon_{q'}\delta^3(\vec{k'}-\vec{q'})(1-n\subs{F}(\beta\cdot k')),
\end{split}
\end{equation*}
one finds
\begin{align}\label{wigner1}
 \Delta W_{+ab}(x,p) =& 2\mu_0 B(x)\int\!\D^3 y \int_0^1\!\!\frac{\D z }{(2\pi)^6} 
 \int \frac{\D^3 \krm\, \D^3 \krm'}{4\varepsilon_k \varepsilon_{k'}}
    \delta^4\left(p-\frac{k+k'}{2} \right)  \\ \nonumber
 &\times \beta n\subs{F}(k)(1-n\subs{F}(k')){\cal S}^z (k,k')_{ab}\, \E^{\I(k-k')(x-y)}\E^{z(k-k')\beta}
\end{align}
with
\begin{equation*}
\begin{split}
{\cal S}^z (k,k')_{ab} \equiv& \frac{\I}{2}(\slashed{k}'+m)
    \gamma^1 \gamma^2 (\slashed{k}+m) .
\end{split}
\end{equation*}
The traces involved in spin polarization read:
\begin{equation*}
\begin{split}
\tr&\left[(\slashed{k}'+m)\gamma^1\gamma^2(\slashed{k}+m)\right]
    =4(k^2 k^{\prime\,1} - k^1 k^{\prime\,2}),\\
\tr&\left[\gamma^\mu\gamma^5(\slashed{k}'+m)\gamma^1\gamma^2(\slashed{k}+m)\right]
    =-4\I m (k_0 + k'_0).
\end{split}
\end{equation*}
Noticing that
\begin{equation*}
    \int\!\D^3 y\, \E^{\I(k-k')(x-y)} = (2\pi)^3 \delta^3(k-k')
\end{equation*}
one see that the first trace is vanishing and hence there is no contribution
from magnetic field to the denominator of the spin polarization formula
(\ref{eq:SpinPolFormula}). The numerator instead is
\begin{equation*}
\begin{split}
\mathcal{N}^z =& \int_\Sigma\!\!\D\Sigma\cdot p\,
        2\mu_0 B(x) \frac{1}{(2\pi)^3} \int \frac{\D^3 \krm\,}{4\varepsilon_k^2}
     4 m \varepsilon_k \delta^4\left(p-k \right)
        \beta n\subs{F}(k)(1-n\subs{F}(k))\\
=& \frac{4m}{(2\pi)^3}\int_\Sigma\!\!\D\Sigma\cdot p\,
        \mu_0 B(x) \frac{\delta(p_0-\varepsilon_p)}{2\varepsilon_p} \beta n\subs{F}(p)(1-n\subs{F}(p))\\
=& \frac{4m}{(2\pi)^3}\int_\Sigma\!\!\D\Sigma\cdot p\,
        \mu_0 B(x) \delta(p^2-m^2)\theta(p_0) \beta n\subs{F}(p)(1-n\subs{F}(p)).
\end{split}
\end{equation*}
The leading order spin polarization resulting from (\ref{eq:RhoB}) is then
\begin{equation*}
S^z(p) = \frac{1}{2m} \frac{\int_{\Sigma} \D \Sigma \cdot p \; \beta n\subs{F} (1 -n\subs{F}) 
 \mu_0\, B\, m  }{\int_{\Sigma} \D \Sigma \cdot p \; n\subs{F}}.
\end{equation*}
As expected this is equal to Eq. (\ref{eq:PolNonRelSubstitution}) in the non-relativistic
limit $\epsilon_p\to m$.

\subsection{Study of global equilibrium spin polarization}
\label{sec:GlobalEqPol}
\begin{figure*}[!tbhp]
   \includegraphics[width=0.5\textwidth]{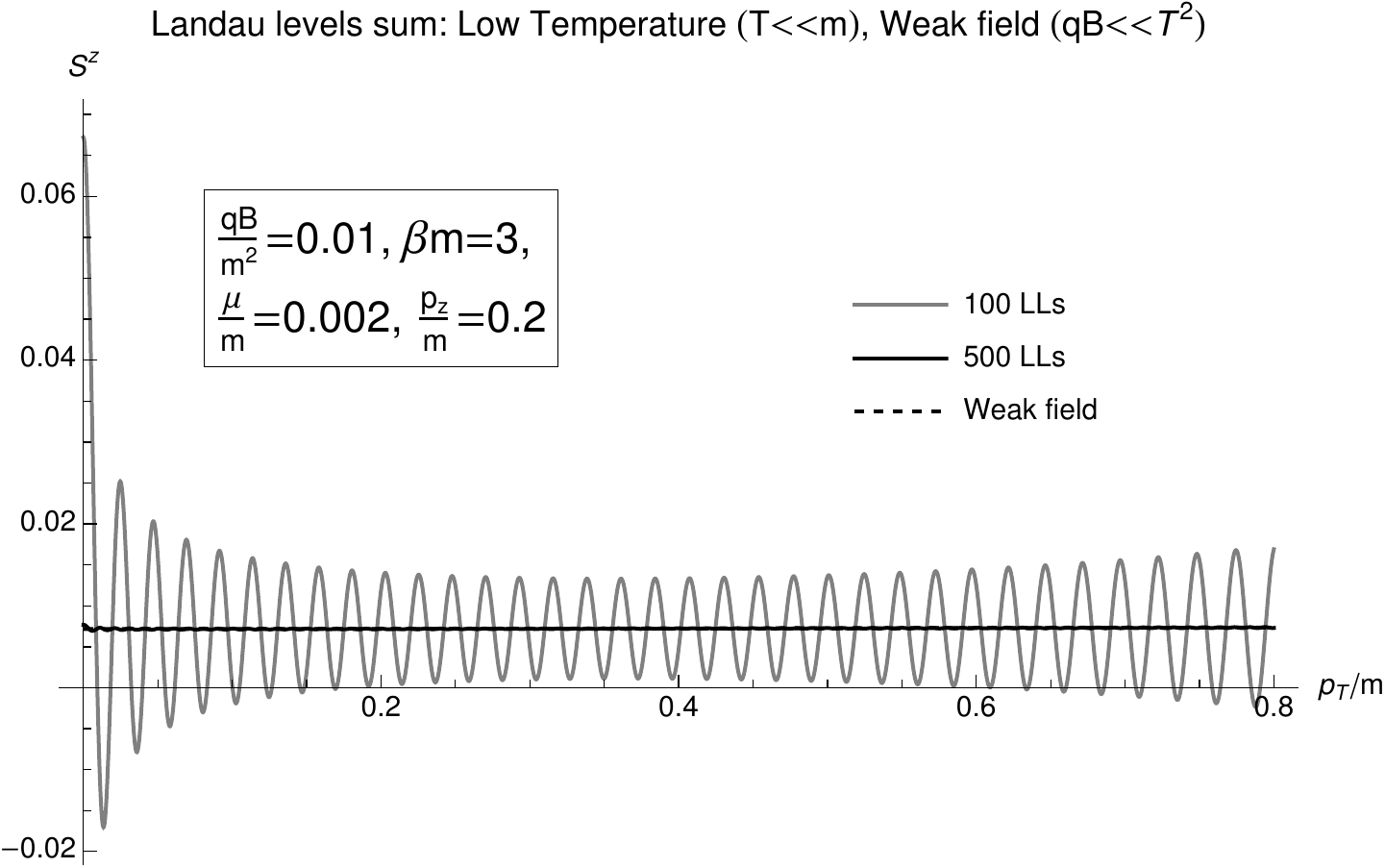}
   \includegraphics[width=0.5\textwidth]{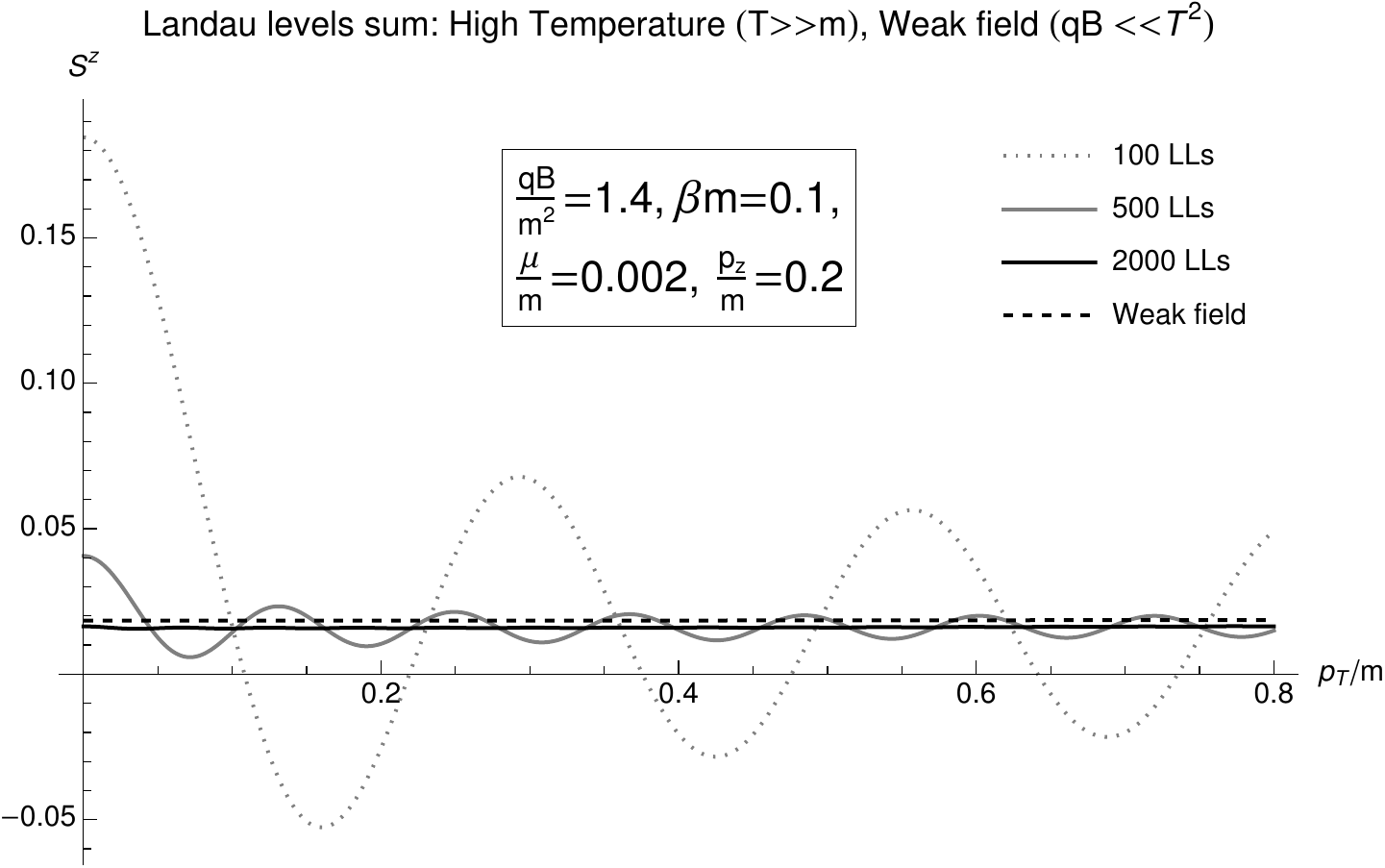}
\caption{The weak field approximation of the spin polarization along the magnetic
    field at global equilibrium (\ref{eq:SpinzBGlobalWeak}) as a function of transverse
    momentum compared with the exact result (\ref{eq:SpinzBGlobalExact}) summed up to 100,
    500 and 2000 Landau levels (LLs).}
\label{fig:LLSums}
\end{figure*}
\begin{figure*}[!tbhp]
   \includegraphics[width=0.5\textwidth]{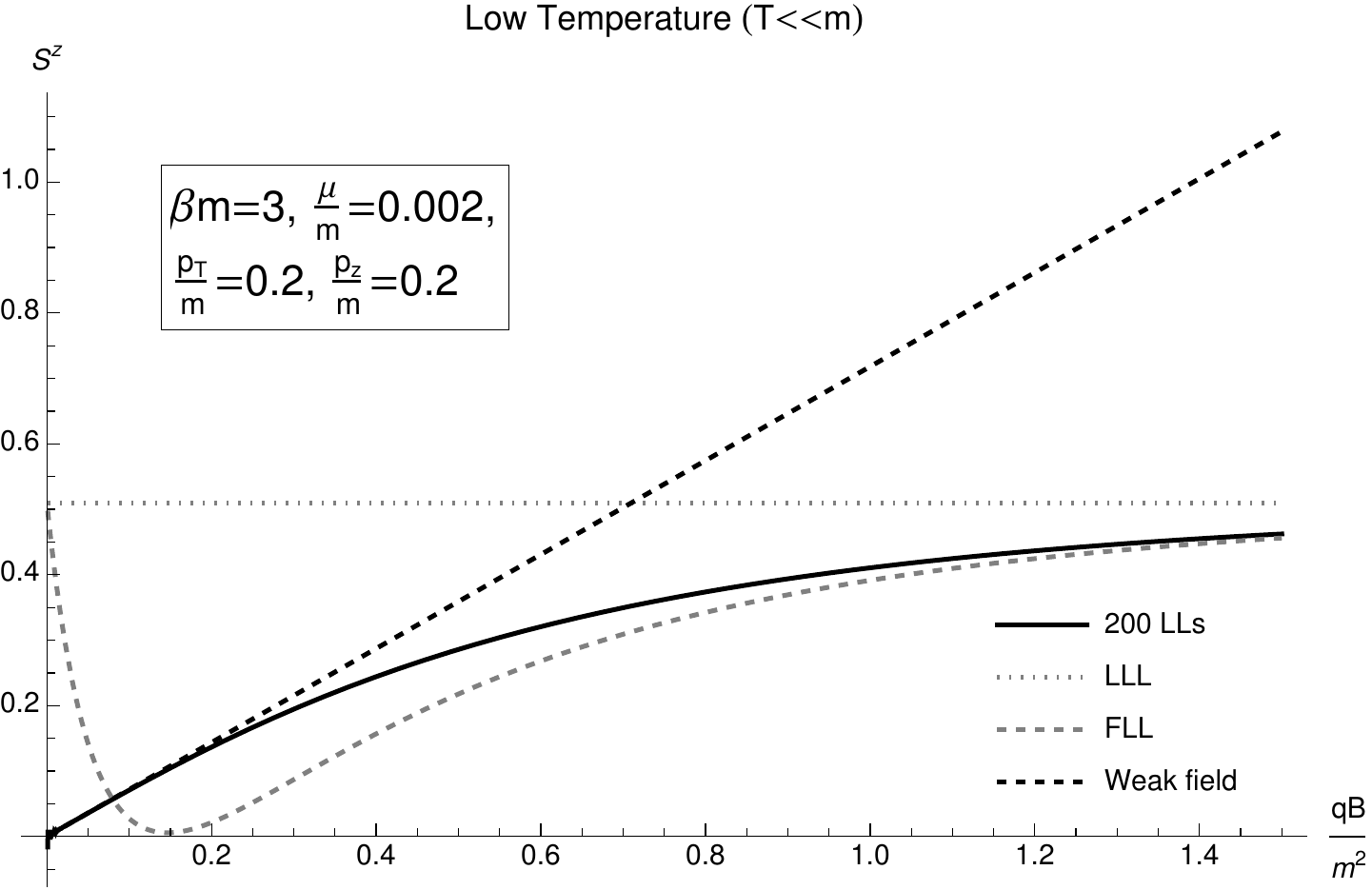}
   \includegraphics[width=0.5\textwidth]{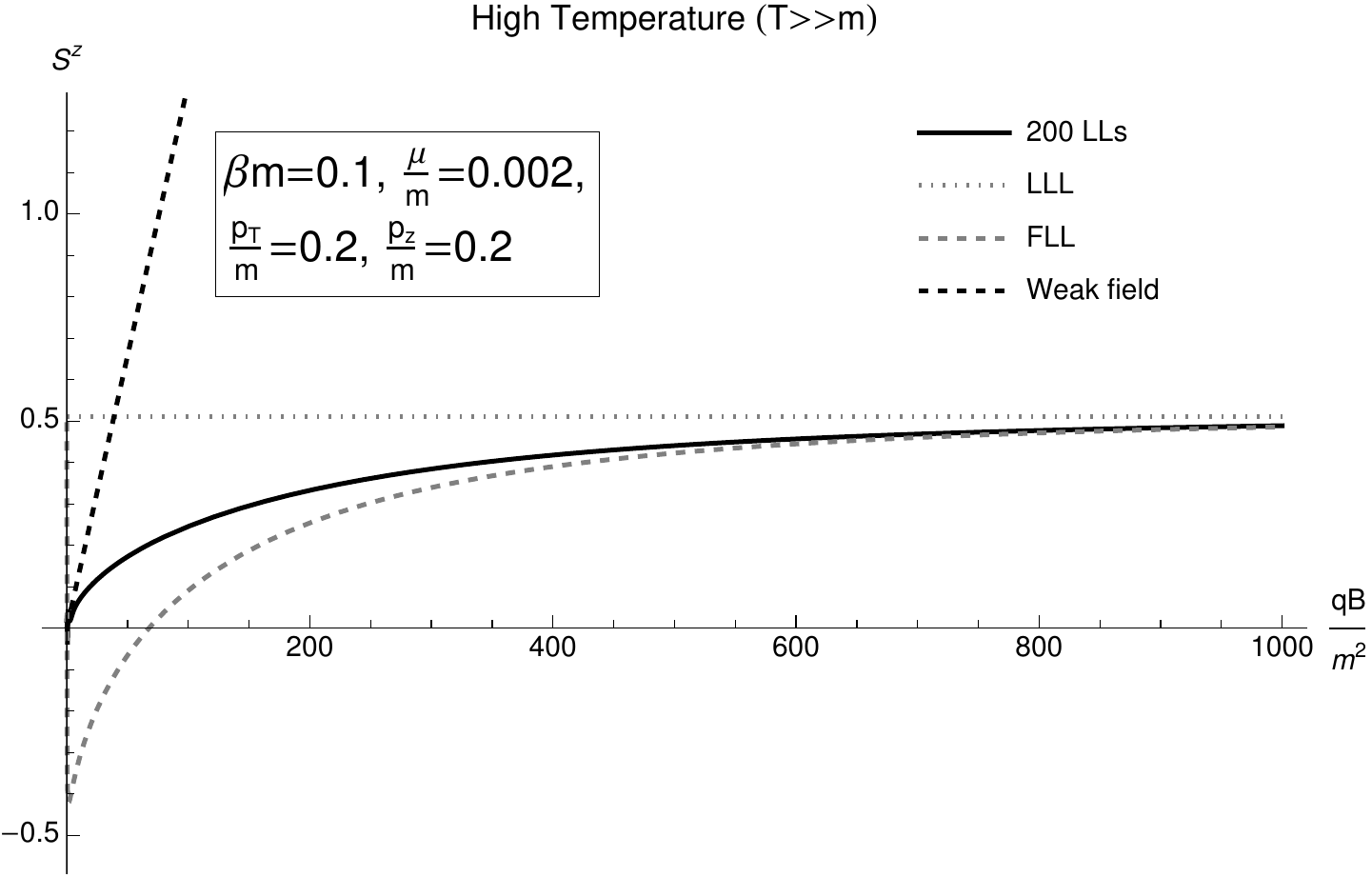}
\caption{The spin polarization along the magnetic field at global equilibrium as a function         of magnetic field intensity at low and high temperatures. The spin polarization is
    evaluated using (\ref{eq:SpinzBGlobalExact}) summing up to $n=0$ (LLL), $n=1$ (FLL)
    and $n=200$ (200 LLs) and using (\ref{eq:SpinzBGlobalWeak}) for the weak field
    approximation.}
\label{fig:qBDependence}
\end{figure*}
Consider a global thermal equilibrium with a constant magnetic field.
Since there is no coordinate dependence, the integral over the hypersurface
can be done in a hypersurface such that its normal vector is $u$:
\begin{equation*}
    \int_{\Sigma} \D \Sigma \cdot p = \varepsilon_p \int \D^3 x  = \varepsilon_p V ,
\end{equation*}
and the spin polarization is obtained from
\begin{equation}
\begin{split}
S^\mu(p)=&\frac{1}{2}\frac{\int_\Sigma \D\Sigma \cdot p\,\mathcal{A}_+^\mu(p)}{\int_\Sigma \D\Sigma\cdot  p\, \mathcal{F}_+(p)}
	=\frac{1}{2}\frac{\mathcal{A}_+^\mu(p)}{\mathcal{F}_+(p)}
	=\frac{1}{2}\frac{\int \D\varepsilon_p\,\mathcal{A}_+^\mu(p)}{\int \D\varepsilon_p\, \mathcal{F}_+(p)}.
\end{split}
\end{equation}
It is convenient to write the exact solution of Wigner function (\ref{eq:ExactWignerF}) and
(\ref{eq:ExactWignerA}) as
\begin{align}
\label{eq:ExactWignerF2}
\mathcal{F}_+(p)=&\sum_{n=0}^\infty \frac{4m}{(2\pi)^3}
    \frac{\delta\left(\varepsilon_p - E_{p_z}^{(n)}\right)}{2\varepsilon_p}\E^{-\xi/2}\\
    &\times n\subs{F}(\beta\cdot p-\zeta)
    (-1)^n \left[ L_n(\xi) -L_{n-1}(\xi) \right],\nonumber\\
\label{eq:ExactWignerA2}
\mathcal{A}_+^\mu(p)  =& \sum_{n=0}^\infty \frac{4 a^\mu}{(2\pi)^3}
    \frac{\delta\left(\varepsilon_p - E_{p_z}^{(n)}\right)}{2\varepsilon_p} \E^{-\xi/2}\\
    &\times n\subs{F}(\beta\cdot p-\zeta)
    (-1)^n \left[ L_n(\xi) + L_{n-1}(\xi) \right],\nonumber
\end{align}
from which it follows that spin polarization at global equilibrium is
\begin{equation}
\begin{split}
S^\mu(p)=&\frac{\sum_n a^\mu n\subs{F}(\beta E_{p_z}^{(n)}-\zeta)
    (-1)^n \left[ L_n(\xi) + L_{n-1}(\xi) \right]}
    {2m\sum_n n\subs{F}(\beta E_{p_z}^{(n)}-\zeta)
        (-1)^n \left[ L_n(\xi) -L_{n-1}(\xi) \right]}.
\end{split}
\end{equation}
In the frame where the fluid is at rest, one has $u^\mu=\delta^\mu_0$, $\vec{B}=B \hat{z}$
and $a^\mu = \left(p^z,0,0,E_{p_z}^{(n)}\right)$. The spin polarization along the direction of
magnetic field is
\begin{equation}
\label{eq:SpinzBGlobalExact}
S^z(p)=\frac{\sum_n E_{p_z}^{(n)}\,n\subs{F}
    (-1)^n \left[ L_n(\xi) + L_{n-1}(\xi) \right]}
    {2m\sum_n n\subs{F}
        (-1)^n \left[ L_n(\xi) -L_{n-1}(\xi) \right]}.
\end{equation}
In the weak field limit, using (\ref{eq:WeakWignerF}) and (\ref{eq:WeakWignerA}), 
\begin{equation}
\label{eq:SpinzBGlobalWeak}
S^z\subs{Weak field}(p)=\frac{1}{4} \beta m \frac{qB}{m^2}
    \left(1-n\subs{F}(\beta\varepsilon_p-\zeta)\right).
\end{equation}
In the very strong field limit the spin polarization reaches the asymptotic value
\begin{equation}
\label{eq:SpinzBLLL}
\lim_{\substack{B \to \infty \\ \sqrt{qB}\gg m,T}} S^z=S^z\subs{LLL}
    = \frac{1}{2} \sqrt{1+\left(\frac{p_z}{m}\right)^2},
\end{equation}
which corresponds to a situation where only the Lowest Landau Level (LLL)
is populated. More accurately, for strong magnetic field the spin polarization
can be approximated by stopping the sums in (\ref{eq:SpinzBGlobalExact}) at the
first landau level (FLL):
\begin{equation}
\label{eq:SpinzBFLL}
\begin{split}
S^z\subs{FLL} =& \frac{1}{2} \sqrt{1+\left(\frac{p_z}{m}\right)^2}
\frac{1+2\frac{p_T^2 -|qB|}{|qB|}\frac{n\subs{F}\left(\beta\sqrt{1+p_z^2+2|qB|}-\zeta\right)}{n\subs{F}\left(\beta\sqrt{1+p_z^2}-\zeta\right)}}{1+2\frac{p_T^2}{|qB|}\frac{\sqrt{1+p_z^2}}{\sqrt{1+p_z^2+2|qB|}}\frac{n\subs{F}\left(\beta\sqrt{1+p_z^2+2|qB|}-\zeta\right)}{n\subs{F}\left(\beta\sqrt{1+p_z^2}-\zeta\right)}}.
\end{split}
\end{equation}

\begin{figure}[!hptb]\centering
    \includegraphics[width=0.6\textwidth]{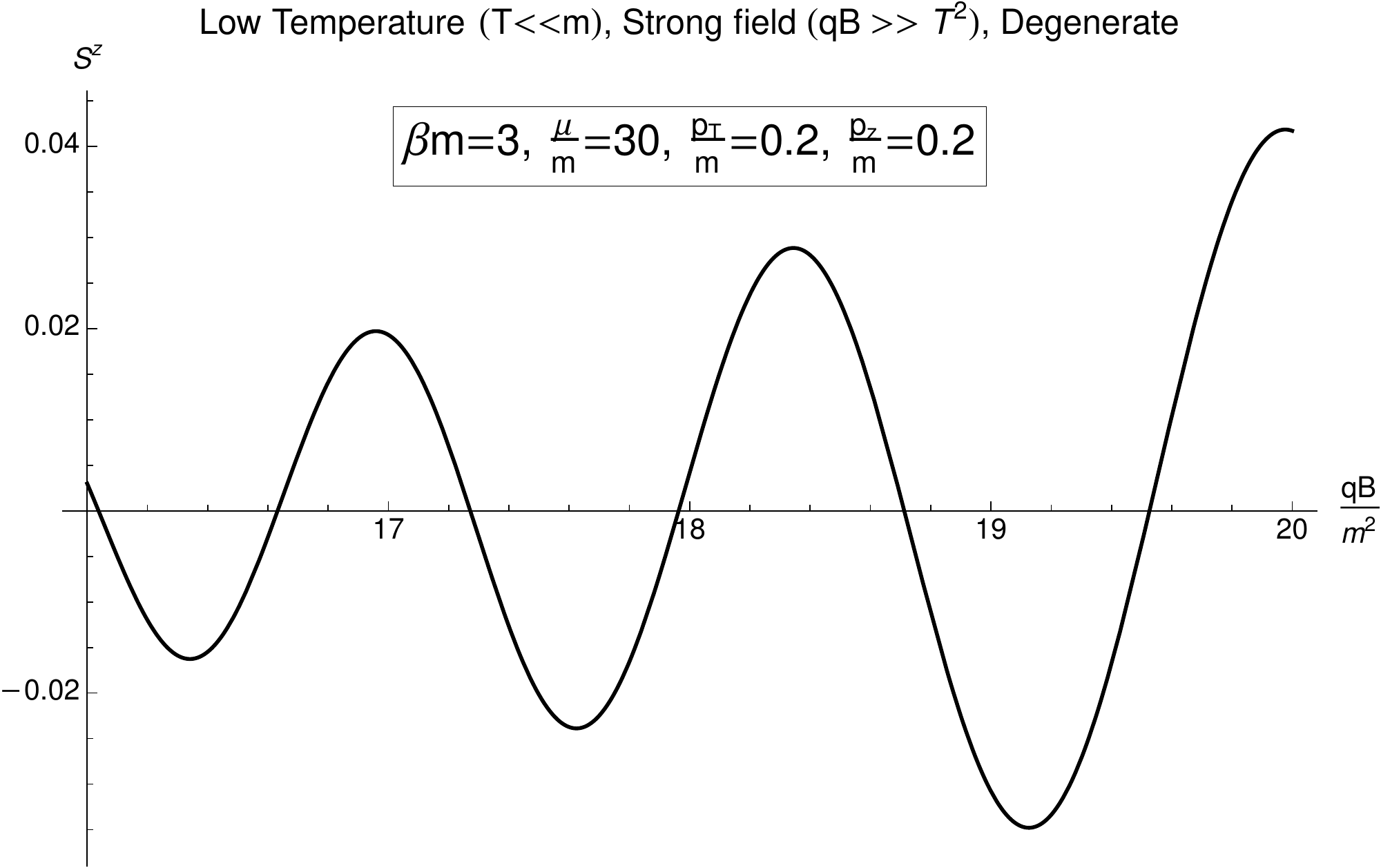}
    \caption{The spin polarization along the magnetic field at global equilibrium
    	at low temperature, strong field and high degeneracy has de Haas - van Alphen oscillations.
    	The line is evaluated using (\ref{eq:SpinzBGlobalExact}) summing up to 500 LLs.}
    \label{fig:dHvH}
\end{figure}

In Fig.~\ref{fig:LLSums} the weak limit approximation (\ref{eq:SpinzBGlobalWeak}) is
evaluated numerically for small values of the magnetic field as a function of transverse
momentum and it is compared with the exact result (\ref{eq:SpinzBGlobalExact}) approximated
by truncating the Landau levels sums at a certain Landau level $N$. The Fig.~\ref{fig:LLSums}
shows that the weak field limit (\ref{eq:SpinzBGlobalWeak}) provides a good approximation of
the exact formula (\ref{eq:SpinzBGlobalExact}). The dependence of the spin polarization on the
magnetic field intensity is studied in Fig.~\ref{fig:qBDependence}. For very large
magnetic field, the spin polarization first approaches the values obtained with
the FLL approximation (\ref{eq:SpinzBFLL}) and then the LLL approximation (\ref{eq:SpinzBLLL}).
Just like the magnetic susceptibility of a highly degenerate system oscillates as the magnetic field
increases according to the de Haas - van Alphen effect~\cite{Landau:1969}, the spin polarization is
expected to have de Haas - van Alphen oscillations as the magnetic field intensity or the
transverse momentum is increased. The oscillations visible in Fig.~\ref{fig:LLSums}
for the lines obtained including few Landau levels
are originated from the Laguerre polynomials in Eq. (\ref{eq:SpinzBGlobalExact}) through
the argument $\xi=2 p_T^2/|qB|$. Those oscillations disappear for a non-degenerate
system such as the one considered in Fig.~\ref{fig:LLSums} and \ref{fig:qBDependence}, but persist for a
degenerate gas in strong magnetic field as Fig.~\ref{fig:dHvH} shows.

%

\subsection{Heavy-ion collisions}
%
\begin{figure}[!hptb]\centering
    \includegraphics[width=0.6\textwidth]{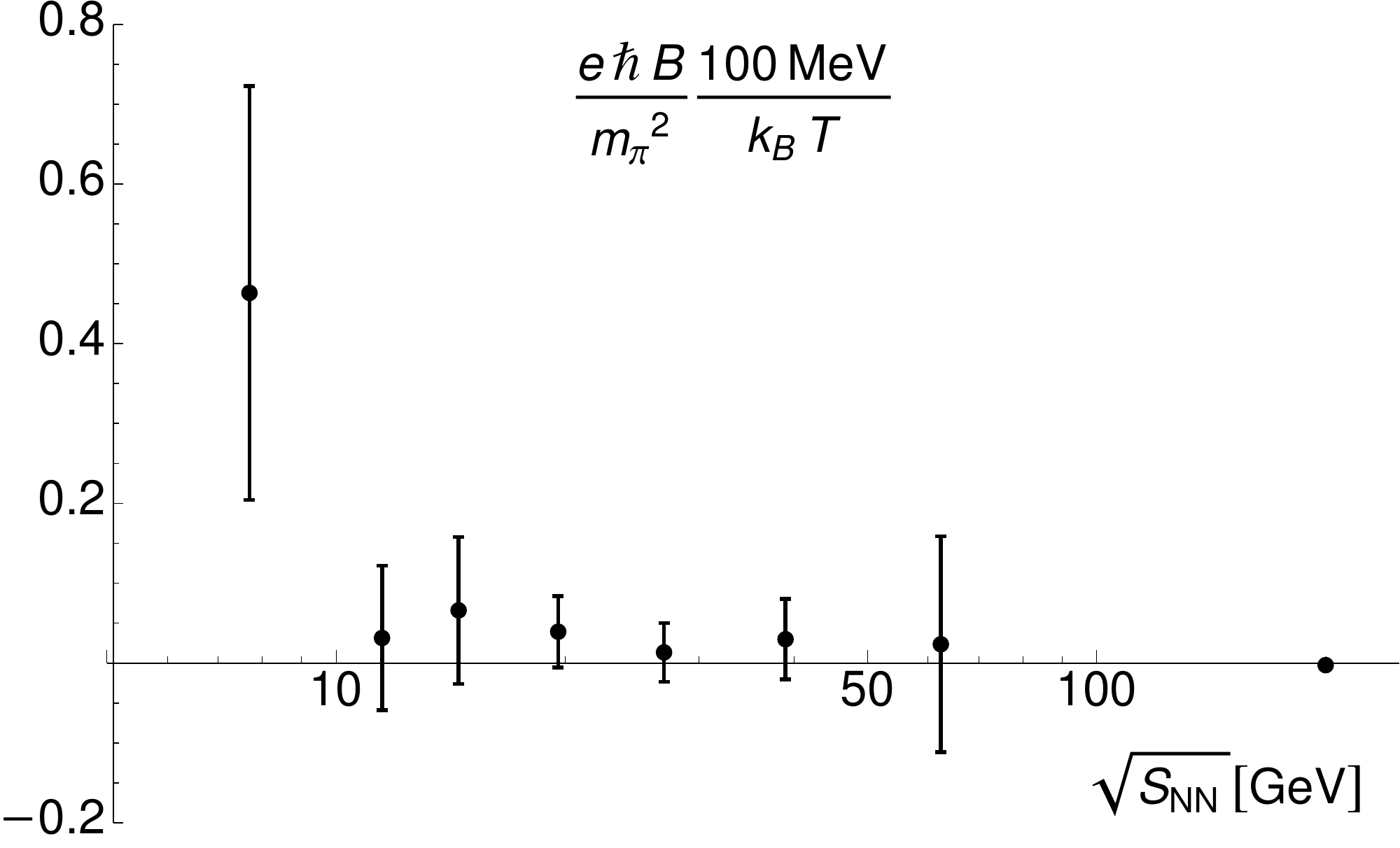}
    \caption{The estimate of magnetic field as a function of energy resulting
    from the difference of global spin polarization \cite{STAR:2017ckg,Adam:2018ivw},
    only the statistical error is shown.}
    \label{fig:eB}
\end{figure}
In heavy-ion collisions the magnetic field and the angular momentum of the plasma
are aligned. The effect of a magnetic field can be revealed in the difference
between the global polarization of $\Lambda$ and $\bar\Lambda$ particles.
The data reported in \cite{STAR:2017ckg,Adam:2018ivw} reveals that this difference
is not consistent with zero only at $\sqrt{S_{NN}}=7.7$ GeV.
Reminding that for spin $1/2$ the polarization vector is related to the spin
vector as
\begin{equation}
    P^\mu = 2 S^\mu,
\end{equation}
adopting the non-relativistic limit of Eq. (\ref{eq:SpinPolFirstOrder}) and
Eq. (\ref{eq:PolWignerB}), the difference in global polarization is estimated as
\begin{equation}
\label{eq:NetDifference}
P_{\bar{\Lambda}}-P_\Lambda \simeq -2 \frac{\mu_\Lambda B}{T}
    -\frac{(\vec{\nabla}\zeta_B \times \vec{v})\cdot \hat{\vec{J}}}{4 m_\Lambda},
\end{equation}
where $\zeta_B$ is the baryonic chemical potential divided by temperature,
and $\mu_\Lambda$ and $m_\Lambda$ are respectively the magnetic moment and the
mass of the $\Lambda$, while $\hat{\vec{J}}$ is the direction of the total
angular momentum. The Fig. \ref{fig:eB} shows the ratio between the magnetic field
and temperature extracted from the experimental data \cite{STAR:2017ckg,Adam:2018ivw}
using the formula above neglecting the contribution from the baryonic chemical potential
and using $\mu_\Lambda=-0.613\mu_N$ \cite{Workman:2022ynf} with $\mu_N$ the nuclear
magnetic moment. 

The spin polarization is only sensitive to fields at the hadronization hypersurface.
It has been showed \cite{Deng:2012pc,McLerran:2013hla,Tuchin:2013apa,Inghirami:2016iru,Guo:2019joy}
that the magnetic field decays very rapidly and that the decay is faster at higher
energies. This estimate shows that the magnetic field is very small at late
stage of the plasma.  The values reported in Fig. \ref{fig:eB} are in agreement with the
calculations of the magnetic field time evolution, possibly overestimating its
value at 7.7 GeV \cite{Xu:2022hql}. An upper bound of the magnetic field can be obtained 
from this data \cite{Muller:2018ibh}. For instance, from the statistical error at 200 GeV,
that is $\Delta(P_{\bar{\Lambda}}-P_\Lambda)=0.052\times 10^{-2}$, one obtains\footnote{The
freeze-out hypersurface is approximately an hypersurface of constant temperature $T\simeq 155$ MeV.}
\begin{equation}
\frac{e\hbar B}{m_\pi^2}\frac{100\text{ MeV}}{k_B T}\leq
    \frac{\Delta(P_{\bar{\Lambda}}-P_\Lambda)}{0.613}\frac{m_p 100\text{ MeV}}{m_\pi^2}
    = 4\times 10^{-3}.
\end{equation}

However, the magnetic field is not the only way the spin polarization can be different
for baryons and anti-baryons. An other effect is showed in Eq. (\ref{eq:NetDifference})
itself: the contribution of the gradient of the baryonic chemical potential. The effect
of an electric field has the same form as the contribution of the gradient of chemical
potential,  but the electric field has been estimated to
be too small. The presence of a chemical potential also changes the thermal distribution
function for $\Lambda$ and $\bar\Lambda$, contributing to this difference. It was also
found that different freeze-out conditions also contribute \cite{Vitiuk:2019rfv}.
Recently all these effects were studied in \cite{Wu:2022mkr,Ryu:2021lnx}. Those
simulations showed that all these effects compete in different directions often
canceling out each other. After all, the estimate of magnetic field presented in
Fig. \ref{fig:eB} is a reliable rough estimate.
In \cite{Ambrus:2020oiw} it was also proposed that the this difference might be caused by the interplay between the axial and helical \cite{Ambrus:2019ayb} currents.

\section{Discussion and conclusion}
I derived the formula for the spin polarization induced by an external magnetic
field in a quantum relativistic framework. The leading order correction for a
weak magnetic field is given by Eq. (\ref{eq:PolWignerB}) and coincides with previous
results \cite{Gao:2012ix,Becattini:2016gvu,Yi:2021ryh} obtained with different methods.
I also derived the full expression given by Eqs. (\ref{eq:SpinPolFormulaB}), (\ref{eq:ExactWignerF})
and (\ref{eq:ExactWignerA}) valid at global equilibrium and at local thermal equilibrium
neglecting the gradients of magnetic field. I studied this expression at global equilibrium
for different magnetic field intensities and I showed that, just like the magnetic susceptibility,
spin polarization in a degenerate gas with strong magnetic field exhibits
de Haas - van Alphen oscillations. These effects are derived assuming that the
system is at local thermal equilibrium, from which it follows that this is a
non-dissipative phenomenon.

While spin polarization was computed for free fields, the framework
used to describe the system is a quantum field theory and interactions can be included
without breaking the quantum properties of the system.
The method used to obtain these results can be extended to higher order in gradients
of the magnetic field provided that the magnetic field changes over macroscopic distances.
As the full effects of thermal vorticity to the distribution function are
difficult to be included without the statistical operator approach \cite{Palermo:2021hlf},
the use of this method could be crucial to study the interplay between electromagnetic
field and thermal vorticity.

Based on these results I reviewed the relativistic Barnett effect and I showed that
the spin polarization induced by vorticity can be obtained with a classical model.
While the model is less accurate (the classical theory fails to reproduce the correct
gyromagnetic moment), it highlights that the effect of vorticity, that is a rotation,
is mainly a classical non-inertial effect. As the classical theory is free of quantum
anomalies, this model also suggests that there is not a tight connection between the
spin polarization of a massive particle and the gravitational anomaly in the axial
current or any other quantum anomaly.

\section*{Acknowledgments}
I am grateful to F. Becattini, A. Palermo and K. Tuchin for valuable discussions.
This work is supported by the US Department of Energy under Grant No. DE-FG02-87ER40371
and DE-SC0023692.

\appendix
\section{Weak field limit}
\label{sec:AppWeak}
In this appendix I derive the results for the weak field approximation of the Wigner
function following \cite{Gorbar:2017awz}. To obtain this limit, one takes advantage of
the following identities
\begin{equation}
\begin{split}
\mathcal{L}_1 = &  \sum_{n=0}^\infty (-1)^n L_n\left( \frac{2 p_T^2}{|qB|} \right) \E^{2 \sigma n |qB|}
 = \frac{\E^{\frac{2p_T^2}{|qB|(1+\E^{-2 \sigma |qB|})}}}{1+\E^{2 \sigma |qB|}},
\end{split}
\end{equation}
and
\begin{equation}
\begin{split}
\mathcal{L}_2 = &  \sum_{n=0}^\infty (-1)^n L_{n-1}\left( \frac{2 p_T^2}{|qB|} \right) \E^{2 \sigma n |qB|}
 =  -\E^{2 \sigma |qB|} \frac{\E^{\frac{2p_T^2}{|qB|(1+\E^{-2 \sigma qB})}}}{1+\E^{2 \sigma |qB|}},
\end{split}
\end{equation}
where $\sigma=\pm$. Combined together the two identities give
\begin{equation}
\label{eq:L1MinusL2}
\begin{split}
\sum_{n=0}^\infty (-1)^n &\left[ L_n\left( \frac{2 p_T^2}{|qB|} \right)
    - L_{n-1}\left( \frac{2 p_T^2}{|qB|} \right)\right] \E^{2 \sigma n |qB|} \\
    =& \mathcal{L}_1 - \mathcal{L}_2 
    = \exp\left[\frac{2p_T^2}{|qB|(1+\E^{-2 \sigma |qB|})}\right],
\end{split}
\end{equation}
and
\begin{equation}
\label{eq:L1PlusL2}
\begin{split}
\sum_{n=0}^\infty (-1)^n &\left[ L_n\left( \frac{2 p_T^2}{|qB|} \right)
    + L_{n-1}\left( \frac{2 p_T^2}{|qB|} \right)\right] \E^{2 \sigma n |qB|} \\
    &= \mathcal{L}_1 + \mathcal{L}_2 
    = \frac{1-\E^{2 \sigma |qB|}}{1+\E^{2 \sigma |qB|}}
    \exp\left[\frac{2p_T^2}{|qB|(1+\E^{-2 \sigma |qB|})}\right],
\end{split}
\end{equation}

Reminding that $\xi=2 p_T^2/|qB|$, consider then the scalar part of the Wigner function
\begin{equation}
\label{eq:AppendixF}
\begin{split}
\mathcal{F}_+(p)=& \sum_{n=0}^\infty \frac{4m}{(2\pi)^3}
    \frac{\delta\left(\varepsilon_p - E_{p_z}^{(n)}\right)}{2\varepsilon_p}
        n\subs{F}(\beta\cdot p-\zeta)\\
    &\times (-1)^n \left[ L_n(\xi) -L_{n-1}(\xi) \right]\E^{-\xi/2}.
\end{split}
\end{equation}
The first part of each term in this series can be expanded as a series
in magnetic field as follow
\begin{equation}
\begin{split}
F\left(p_z^2+2n|qB| \right) = &
\frac{4m}{(2\pi)^3} \frac{\delta\left(\varepsilon_p - E_{p_z}^{(n)}\right)}{2\varepsilon_p}
    n\subs{F}(\beta\cdot p-\zeta) \\ 
= & \sum_{l=0}^\infty (2n |qB|)^l \frac{F^{(l)}(p_z^2)}{l!},
\end{split}
\end{equation}
where I denoted
\begin{equation}
    F^{(l)}(z) = \frac{\D^l}{\D z^l} F(z).
\end{equation}
Since each coefficient of this expansion can be obtained as the limit
\begin{equation}
\begin{split}
F\left(p_z^2+2n|qB| \right) = & \lim_{s \to 0^-}
\sum_{l=0}^\infty \left(\frac{\D^l}{\D s^l} \E^{2 s |qB| n}\right)
\frac{F^{(l)}(p_z^2)}{l!},
\end{split}
\end{equation}
one obtains
\begin{equation}
\begin{split}
\mathcal{F}_+(p)=& \lim_{s \to 0^-} \sum_{n=0}^\infty \sum_{l=0}^\infty
    \left(\frac{\D^l}{\D s^l} \E^{2 s |qB| n}\right) \frac{F^{(l)}(p_z^2)}{l!}\\
    &\times (-1)^n \left[ L_n(\xi) -L_{n-1}(\xi) \right]\E^{-\xi/2}.
\end{split}
\end{equation}
Exchanging the order of the summations, one finds the identity (\ref{eq:L1MinusL2})
and
\begin{equation}
\begin{split}
\mathcal{F}_+(p)=& \lim_{s \to 0^-}  \sum_{l=0}^\infty \frac{F^{(l)}(p_z^2)}{l!}\E^{-\xi/2}\\
    &\times \frac{\D^l}{\D s^l} \sum_{n=0}^\infty (-1)^n \left[ L_n(\xi) -L_{n-1}(\xi) \right]
    \E^{2 s |qB| n}\\
= & \lim_{s \to 0^-} \sum_{l=0}^\infty \frac{F^{(l)}(p_z^2)}{l!}\E^{-\xi/2}
    \frac{\D^l}{\D s^l} \E^{\frac{2p_T^2}{|qB|(1+\E^{-2 s |qB|})}}.
\end{split}
\end{equation}
Keeping terms up to $|qB|^3$, the scalar part of Wigner function is approximated by
\begin{equation}
\begin{split}
\mathcal{F}_+(p)\simeq &\lim_{s \to 0^-}  \sum_{l=0}^\infty \frac{F^{(l)}(p_z^2)}{l!}\E^{-\xi/2}
    \frac{\D^l}{\D s^l}\left[ \E^{s p_T^2-\tfrac{1}{3}p_T^2 |qB|^2 s^3}
        +\mathcal{O}(|qB|^3)\right]\\
\simeq &  \sum_{l=0}^\infty \frac{F^{(l)}(p_z^2)}{l!}p_T^{2l} + \mathcal{O}(|qB|^2)\\
= & F(p_z^2+p_T^2) + \mathcal{O}(|qB|^2) \\
= & \frac{4m}{(2\pi)^3} \delta\left(\varepsilon_p - E_p\right)
    \frac{n\subs{F}(\beta\cdot p-\zeta)}{2\varepsilon_p}+ \mathcal{O}(|qB|^2)\\
= & \frac{4m}{(2\pi)^3} \delta\left(\varepsilon_p^2 - E_p^2\right)\theta(\varepsilon_p)
    n\subs{F}(\beta\cdot p-\mu)\\
= & \frac{4m}{(2\pi)^3} \delta\left(p^2 - m^2\right)\theta(\varepsilon_p)
    n\subs{F}(\beta\cdot p-\mu),
\end{split}
\end{equation}
where I used the properties of the Dirac delta and I denoted
$E_p = \sqrt{m^2 + \vec{p}^2}$.

The weak field limit of the axial part of the Wigner function is obtained
in a similar fashion. Define the function
\begin{equation}
A^\mu (p_z^2+2n|qB|)= \frac{4 a^\mu\theta(\varepsilon_p)}{(2\pi)^3}
    \delta\left(\varepsilon_p^2 - {E_{p_z}^{(n)}}^2\right)
     n\subs{F}(\beta\cdot p-\zeta),
\end{equation}
such that as shown for the scalar part
\begin{equation}
\begin{split}
\mathcal{A}_+^\mu(p)  =& \sum_{n=0}^\infty \frac{4 a^\mu\theta(\varepsilon_p)}{(2\pi)^3}
    \delta\left(\varepsilon_p^2 - {E_{p_z}^{(n)}}^2\right)
        n\subs{F}(\beta\cdot p-\zeta)\\
    &\times (-1)^n \left[ L_n(\xi) + L_{n-1}(\xi) \right] \E^{-\xi/2}\\
=& \lim_{s \to 0^-} \sum_{n=0}^\infty \sum_{l=0}^\infty
    \left(\frac{\D^l}{\D s^l} \E^{2 s |qB| n}\right) \frac{A^{\mu(l)}(p_z^2)}{l!}\\
    &\times (-1)^n \left[ L_n(\xi) +L_{n-1}(\xi) \right]\E^{-\xi/2}.
\end{split}
\end{equation}
Exchanging the order of the summations and using (\ref{eq:L1PlusL2}),
the axial part becomes
\begin{equation}
\begin{split}
\mathcal{A}^\mu_+(p)= & \lim_{s \to 0^-} \sum_{l=0}^\infty
    \frac{A^{\mu(l)}(p_z^2)}{l!}\E^{-\xi/2}\\
    &\times\frac{\D^l}{\D s^l}\frac{1-\E^{2 s |qB|}}{1+\E^{2 s |qB|}}
     \E^{\frac{2p_T^2}{|qB|(1+\E^{-2 s |qB|})}}.
\end{split}
\end{equation}
Up to the first order in $|qB|$, the expansion of the axial part is
\begin{equation}
\begin{split}
\mathcal{A}^\mu_+(p)\simeq & \lim_{s \to 0^-} -\sum_{l=0}^\infty
    \frac{A^{\mu(l)}(p_z^2)}{l!}
    \frac{\D^l}{\D s^l}\left[\E^{p_T^2 s}|qB| + \cdots \right]\\
= & -\sum_{l=0}^\infty \frac{A^{\mu(l)}(p_z^2)}{(l-1)!}p_T^{2(l-1)}|qB|
    + \mathcal{O}\left(|qB|^2\right).
\end{split}
\end{equation}
Noticing that the series is the derivative of the function $A^\mu$:
\begin{equation}
\mathcal{A}^\mu_+(p)= -|qB|\frac{\D}{\D p_T^2} A^{\mu}(p_z^2+p_T^2)
    + \mathcal{O}\left(|qB|^2\right),
\end{equation}
that the derivative of
\begin{equation}
A^{\mu}(p_z^2+p_T^2) =\frac{4 a^\mu\theta(\varepsilon_p)}{(2\pi)^3}
    \delta\left(\varepsilon_p^2 - {E_p}^2\right)
     n\subs{F}(\beta\cdot p-\zeta)
\end{equation}
acts only on the delta function
\begin{equation}
\begin{split}
\frac{\D}{\D p_T^2} \delta\left(\varepsilon_p^2 - {E_p}^2\right)
    =& \frac{\D E_p}{\D p_T^2} \frac{\D}{\D E_p} \delta\left(\varepsilon_p^2 - {E_p}^2\right)\\
=& \frac{1}{2 E_p}\frac{\D}{\D E_p} \delta\left(\varepsilon_p^2 - {E_p}^2\right)
    =-\delta'\left(p^2 - m^2\right),
\end{split}
\end{equation}
one obtains
\begin{equation}
\begin{split}
\mathcal{A}_+^\mu(p) =& \frac{4|qB|a^\mu}{(2\pi)^3}n\subs{F}(\beta\cdot p-\zeta)
     \theta(\varepsilon_p)\delta'(p^2-m^2).
\end{split}
\end{equation}
%

\begin{thebibliography}{50}
\expandafter\ifx\csname url\endcsname\relax
  \def\url#1{\texttt{#1}}\fi
\expandafter\ifx\csname urlprefix\endcsname\relax\def\urlprefix{URL }\fi
\expandafter\ifx\csname href\endcsname\relax
  \def\href#1#2{#2} \def\path#1{#1}\fi

\bibitem{Tuchin:2013ie}
K.~Tuchin, {Particle production in strong electromagnetic fields in
  relativistic heavy-ion collisions}, Adv. High Energy Phys. 2013 (2013)
  490495.
\newblock \href {http://arxiv.org/abs/1301.0099} {\path{arXiv:1301.0099}},
  \href {https://doi.org/10.1155/2013/490495} {\path{doi:10.1155/2013/490495}}.

\bibitem{Becattini:2007sr}
F.~Becattini, F.~Piccinini, J.~Rizzo, {Angular momentum conservation in heavy
  ion collisions at very high energy}, Phys. Rev. C 77 (2008) 024906.
\newblock \href {http://arxiv.org/abs/0711.1253} {\path{arXiv:0711.1253}},
  \href {https://doi.org/10.1103/PhysRevC.77.024906}
  {\path{doi:10.1103/PhysRevC.77.024906}}.

\bibitem{STAR:2017ckg}
L.~Adamczyk, et~al., {Global $\Lambda$ hyperon polarization in nuclear
  collisions: evidence for the most vortical fluid}, Nature 548 (2017) 62--65.
\newblock \href {http://arxiv.org/abs/1701.06657} {\path{arXiv:1701.06657}},
  \href {https://doi.org/10.1038/nature23004} {\path{doi:10.1038/nature23004}}.

\bibitem{Adam:2018ivw}
J.~Adam, et~al., {Global polarization of $\Lambda$ hyperons in Au+Au collisions
  at $\sqrt{s_{_{NN}}}$ = 200 GeV}, Phys. Rev. C 98 (2018) 014910.
\newblock \href {http://arxiv.org/abs/1805.04400} {\path{arXiv:1805.04400}},
  \href {https://doi.org/10.1103/PhysRevC.98.014910}
  {\path{doi:10.1103/PhysRevC.98.014910}}.

\bibitem{Adam:2019srw}
J.~Adam, et~al., {Polarization of $\Lambda$ ($\bar{\Lambda}$) hyperons along
  the beam direction in Au+Au collisions at $\sqrt{s_{_{NN}}}$ = 200 GeV},
  Phys. Rev. Lett. 123~(13) (2019) 132301.
\newblock \href {http://arxiv.org/abs/1905.11917} {\path{arXiv:1905.11917}},
  \href {https://doi.org/10.1103/PhysRevLett.123.132301}
  {\path{doi:10.1103/PhysRevLett.123.132301}}.

\bibitem{ALICE:2019aid}
S.~Acharya, et~al., {Evidence of Spin-Orbital Angular Momentum Interactions in
  Relativistic Heavy-Ion Collisions}, Phys. Rev. Lett. 125~(1) (2020) 012301.
\newblock \href {http://arxiv.org/abs/1910.14408} {\path{arXiv:1910.14408}},
  \href {https://doi.org/10.1103/PhysRevLett.125.012301}
  {\path{doi:10.1103/PhysRevLett.125.012301}}.

\bibitem{STAR:2020xbm}
J.~Adam, et~al., {Global Polarization of $\Xi$ and $\Omega$ Hyperons in Au+Au
  Collisions at $\sqrt {s_{NN}}$ = 200 GeV}, Phys. Rev. Lett. 126~(16) (2021)
  162301.
\newblock \href {http://arxiv.org/abs/2012.13601} {\path{arXiv:2012.13601}},
  \href {https://doi.org/10.1103/PhysRevLett.126.162301}
  {\path{doi:10.1103/PhysRevLett.126.162301}}.

\bibitem{STAR:2021beb}
M.~S. Abdallah, et~al., {Global $\Lambda$-hyperon polarization in Au+Au
  collisions at $\sqrt {s_{NN}}$=3~GeV}, Phys. Rev. C 104~(6) (2021) L061901.
\newblock \href {http://arxiv.org/abs/2108.00044} {\path{arXiv:2108.00044}},
  \href {https://doi.org/10.1103/PhysRevC.104.L061901}
  {\path{doi:10.1103/PhysRevC.104.L061901}}.

\bibitem{Becattini:2020ngo}
F.~Becattini, M.~A. Lisa, {Polarization and Vorticity in the Quark Gluon
  Plasma}, Annual Review of Nuclear and Particle Science 70~(1) (2020)
  395--423.
\newblock \href {http://arxiv.org/abs/2003.03640} {\path{arXiv:2003.03640}},
  \href {https://doi.org/10.1146/annurev-nucl-021920-095245}
  {\path{doi:10.1146/annurev-nucl-021920-095245}}.

\bibitem{Buzzegoli:2022kyj}
M.~Buzzegoli, {Polarization in heavy ion collisions: A theoretical review}, EPJ
  Web Conf. 276 (2023) 01011.
\newblock \href {http://arxiv.org/abs/2209.06879} {\path{arXiv:2209.06879}},
  \href {https://doi.org/10.1051/epjconf/202327601011}
  {\path{doi:10.1051/epjconf/202327601011}}.

\bibitem{Barnett:1915}
S.~J. Barnett, {Magnetization by Rotation}, Physical Review 6 (1915) 239--270.
\newblock \href {https://doi.org/10.1103/PhysRev.6.239}
  {\path{doi:10.1103/PhysRev.6.239}}.

\bibitem{Barnett:1935}
S.~J. Barnett, {Gyromagnetic and Electron-Inertia Effects}, Rev. Mod. Phys. 7
  (1935) 129--166.
\newblock \href {https://doi.org/10.1103/RevModPhys.7.129}
  {\path{doi:10.1103/RevModPhys.7.129}}.

\bibitem{Koide:2012kx}
T.~Koide, {Spin-electromagnetic hydrodynamics and magnetization induced by
  spin-magnetic interaction}, Phys. Rev. C 87~(3) (2013) 034902.
\newblock \href {http://arxiv.org/abs/1206.1976} {\path{arXiv:1206.1976}},
  \href {https://doi.org/10.1103/PhysRevC.87.034902}
  {\path{doi:10.1103/PhysRevC.87.034902}}.

\bibitem{Singh:2022ltu}
R.~Singh, M.~Shokri, S.~M. A.~T. Mehr, {Relativistic magnetohydrodynamics with
  spin}, arXiv preprint (2 2022).
\newblock \href {http://arxiv.org/abs/2202.11504} {\path{arXiv:2202.11504}}.

\bibitem{Bhadury:2022ulr}
S.~Bhadury, W.~Florkowski, A.~Jaiswal, A.~Kumar, R.~Ryblewski, {Relativistic
  Spin Magnetohydrodynamics}, Phys. Rev. Lett. 129~(19) (2022) 192301.
\newblock \href {http://arxiv.org/abs/2204.01357} {\path{arXiv:2204.01357}},
  \href {https://doi.org/10.1103/PhysRevLett.129.192301}
  {\path{doi:10.1103/PhysRevLett.129.192301}}.

\bibitem{Gao:2012ix}
J.-H. Gao, Z.-T. Liang, S.~Pu, Q.~Wang, X.-N. Wang, {Chiral Anomaly and Local
  Polarization Effect from Quantum Kinetic Approach}, Phys. Rev. Lett. 109
  (2012) 232301.
\newblock \href {http://arxiv.org/abs/1203.0725} {\path{arXiv:1203.0725}},
  \href {https://doi.org/10.1103/PhysRevLett.109.232301}
  {\path{doi:10.1103/PhysRevLett.109.232301}}.

\bibitem{Becattini:2016gvu}
F.~Becattini, I.~Karpenko, M.~Lisa, I.~Upsal, S.~Voloshin, {Global hyperon
  polarization at local thermodynamic equilibrium with vorticity, magnetic
  field and feed-down}, Phys. Rev. C 95~(5) (2017) 054902.
\newblock \href {http://arxiv.org/abs/1610.02506} {\path{arXiv:1610.02506}},
  \href {https://doi.org/10.1103/PhysRevC.95.054902}
  {\path{doi:10.1103/PhysRevC.95.054902}}.

\bibitem{Yi:2021ryh}
C.~Yi, S.~Pu, D.-L. Yang, {Reexamination of local spin polarization beyond
  global equilibrium in relativistic heavy ion collisions}, Phys. Rev. C
  104~(6) (2021) 064901.
\newblock \href {http://arxiv.org/abs/2106.00238} {\path{arXiv:2106.00238}},
  \href {https://doi.org/10.1103/PhysRevC.104.064901}
  {\path{doi:10.1103/PhysRevC.104.064901}}.

\bibitem{Guo:2019joy}
Y.~Guo, S.~Shi, S.~Feng, J.~Liao, {Magnetic Field Induced Polarization
  Difference between Hyperons and Anti-hyperons}, Phys. Lett. B 798 (2019)
  134929.
\newblock \href {http://arxiv.org/abs/1905.12613} {\path{arXiv:1905.12613}},
  \href {https://doi.org/10.1016/j.physletb.2019.134929}
  {\path{doi:10.1016/j.physletb.2019.134929}}.

\bibitem{Xu:2022hql}
K.~Xu, F.~Lin, A.~Huang, M.~Huang,
  {\ensuremath{\Lambda}/\ensuremath{\Lambda}\textasciimacron{} polarization and
  splitting induced by rotation and magnetic field}, Phys. Rev. D 106~(7)
  (2022) L071502.
\newblock \href {http://arxiv.org/abs/2205.02420} {\path{arXiv:2205.02420}},
  \href {https://doi.org/10.1103/PhysRevD.106.L071502}
  {\path{doi:10.1103/PhysRevD.106.L071502}}.

\bibitem{Ryu:2021lnx}
S.~Ryu, V.~Jupic, C.~Shen, {Probing early-time longitudinal dynamics with the
  \ensuremath{\Lambda} hyperon's spin polarization in relativistic heavy-ion
  collisions}, Phys. Rev. C 104~(5) (2021) 054908.
\newblock \href {http://arxiv.org/abs/2106.08125} {\path{arXiv:2106.08125}},
  \href {https://doi.org/10.1103/PhysRevC.104.054908}
  {\path{doi:10.1103/PhysRevC.104.054908}}.

\bibitem{Wu:2022mkr}
X.-Y. Wu, C.~Yi, G.-Y. Qin, S.~Pu, {Local and global polarization of $\Lambda$
  hyperons across RHIC-BES energies: the roles of spin hall effect, initial
  condition and baryon diffusion}, Phys. Rev. C 105 (2022) 064909.
\newblock \href {http://arxiv.org/abs/2204.02218} {\path{arXiv:2204.02218}},
  \href {https://doi.org/10.1103/PhysRevC.105.064909}
  {\path{doi:10.1103/PhysRevC.105.064909}}.

\bibitem{Muller:2018ibh}
B.~M\"uller, A.~Sch\"afer, {Chiral magnetic effect and an experimental bound on
  the late time magnetic field strength}, Phys. Rev. D 98~(7) (2018) 071902.
\newblock \href {http://arxiv.org/abs/1806.10907} {\path{arXiv:1806.10907}},
  \href {https://doi.org/10.1103/PhysRevD.98.071902}
  {\path{doi:10.1103/PhysRevD.98.071902}}.

\bibitem{Adams:2012th}
A.~Adams, L.~D. Carr, T.~Sch\"afer, P.~Steinberg, J.~E. Thomas, {Strongly
  Correlated Quantum Fluids: Ultracold Quantum Gases, Quantum Chromodynamic
  Plasmas, and Holographic Duality}, New J. Phys. 14 (2012) 115009.
\newblock \href {http://arxiv.org/abs/1205.5180} {\path{arXiv:1205.5180}},
  \href {https://doi.org/10.1088/1367-2630/14/11/115009}
  {\path{doi:10.1088/1367-2630/14/11/115009}}.

\bibitem{Zubarev:1966}
D.~N. Zubarev, {A statistical operator for non stationary processes}, Sov.
  Phys. Doklady 10 (1966) 850.

\bibitem{Zubarev:1979}
D.~N. Zubarev, A.~V. Prozorkevich, S.~A. Smolyanskii,
  \href{http://dx.doi.org/10.1007/BF01032069}{Derivation of nonlinear
  generalized equations of quantum relativistic hydrodynamics}, Theoretical and
  Mathematical Physics 40~(3) (1979) 821--831.
\newblock \href {https://doi.org/10.1007/BF01032069}
  {\path{doi:10.1007/BF01032069}}.
\newline\urlprefix\url{http://dx.doi.org/10.1007/BF01032069}

\bibitem{vanWeert1982}
C.~G. van Weert, Maximum entropy principle and relativistic hydrodynamics,
  Annals of Physics 140 (1982) 133--162.
\newblock \href {https://doi.org/10.1016/0003-4916(82)90338-4}
  {\path{doi:10.1016/0003-4916(82)90338-4}}.

\bibitem{Zubarev:1989su}
D.~N. Zubarev, M.~V. Tokarchuk, {Nonequilibrium Thermo Field Dynamics and the
  Method of the Nonequilibrium Statistical Operator}, Teor. Mat. Fiz. 88N2
  (1991) 286--310.
\newblock \href {https://doi.org/10.1007/BF01019114}
  {\path{doi:10.1007/BF01019114}}.

\bibitem{Becattini:2014yxa}
F.~Becattini, L.~Bucciantini, E.~Grossi, L.~Tinti, {Local thermodynamical
  equilibrium and the beta frame for a quantum relativistic fluid}, Eur. Phys.
  J. C 75~(5) (2015) 191.
\newblock \href {http://arxiv.org/abs/1403.6265} {\path{arXiv:1403.6265}},
  \href {https://doi.org/10.1140/epjc/s10052-015-3384-y}
  {\path{doi:10.1140/epjc/s10052-015-3384-y}}.

\bibitem{Buzzegoli:2018wpy}
M.~Buzzegoli, F.~Becattini, {General thermodynamic equilibrium with axial
  chemical potential for the free Dirac field}, JHEP 12~(2018) 002,
  [Erratum: JHEP03,045(2022)].
\newblock \href {http://arxiv.org/abs/1807.02071} {\path{arXiv:1807.02071}},
  \href {https://doi.org/10.1007/JHEP12(2018)002}
  {\path{doi:10.1007/JHEP12(2018)002}}.

\bibitem{Becattini:2019dxo}
F.~Becattini, M.~Buzzegoli, E.~Grossi, {Reworking the Zubarev's approach to
  non-equilibrium quantum statistical mechanics}, Particles 2~(2) (2019)
  197--207.
\newblock \href {http://arxiv.org/abs/1902.01089} {\path{arXiv:1902.01089}},
  \href {https://doi.org/10.3390/particles2020014}
  {\path{doi:10.3390/particles2020014}}.

\bibitem{Buzzegoli:2020ycf}
M.~Buzzegoli, {Thermodynamic equilibrium of massless fermions with vorticity,
  chirality and electromagnetic field}, Lect. Notes Phys. 987 (2021) 53--93.
\newblock \href {http://arxiv.org/abs/2011.09974} {\path{arXiv:2011.09974}},
  \href {https://doi.org/10.1007/978-3-030-71427-7_3}
  {\path{doi:10.1007/978-3-030-71427-7_3}}.

\bibitem{Sheng:2017lfu}
X.-l. Sheng, D.~H. Rischke, D.~Vasak, Q.~Wang, {Wigner functions for fermions
  in strong magnetic fields}, Eur. Phys. J. A 54~(2) (2018) 21.
\newblock \href {http://arxiv.org/abs/1707.01388} {\path{arXiv:1707.01388}},
  \href {https://doi.org/10.1140/epja/i2018-12414-9}
  {\path{doi:10.1140/epja/i2018-12414-9}}.

\bibitem{Gorbar:2017awz}
E.~V. Gorbar, V.~A. Miransky, I.~A. Shovkovy, P.~O. Sukhachov, {Wigner function
  and kinetic phenomena for chiral plasma in a strong magnetic field}, JHEP
  08~(2017) 103.
\newblock \href {http://arxiv.org/abs/1707.01105} {\path{arXiv:1707.01105}},
  \href {https://doi.org/10.1007/JHEP08(2017)103}
  {\path{doi:10.1007/JHEP08(2017)103}}.

\bibitem{Matsuo2015}
M.~Matsuo, J.~Ieda, S.~Maekawa, {Mechanical generation of spin current},
  Frontiers in Physics 3~(July) (2015) 1--10.
\newblock \href {https://doi.org/10.3389/fphy.2015.00054}
  {\path{doi:10.3389/fphy.2015.00054}}.

\bibitem{deOliveira:1962apw}
C.~G. de~Oliveira, J.~Tiomno, {Representations of Dirac equation in general
  relativity}, Nuovo Cim. 24~(4) (1962) 672--687.
\newblock \href {https://doi.org/10.1007/BF02816716}
  {\path{doi:10.1007/BF02816716}}.

\bibitem{Hehl:1990nf}
F.~W. Hehl, W.-T. Ni, {Inertial effects of a Dirac particle}, Phys. Rev. D42
  (1990) 2045--2048.
\newblock \href {https://doi.org/10.1103/PhysRevD.42.2045}
  {\path{doi:10.1103/PhysRevD.42.2045}}.

\bibitem{Becattini:2013fla}
F.~Becattini, V.~Chandra, L.~Del~Zanna, E.~Grossi, {Relativistic distribution
  function for particles with spin at local thermodynamical equilibrium},
  Annals Phys. 338 (2013) 32--49.
\newblock \href {http://arxiv.org/abs/1303.3431} {\path{arXiv:1303.3431}},
  \href {https://doi.org/10.1016/j.aop.2013.07.004}
  {\path{doi:10.1016/j.aop.2013.07.004}}.

\bibitem{Buzzegoli:2017cqy}
M.~Buzzegoli, E.~Grossi, F.~Becattini, {General equilibrium second-order
  hydrodynamic coefficients for free quantum fields}, JHEP 10~(2017)
  091, [Erratum: JHEP07,119(2018)].
\newblock \href {http://arxiv.org/abs/1704.02808} {\path{arXiv:1704.02808}},
  \href {https://doi.org/10.1007/JHEP10(2017)091}
  {\path{doi:10.1007/JHEP10(2017)091}}.

\bibitem{Buzzegoli:2021jeh}
M.~Buzzegoli, D.~E. Kharzeev, {Anomalous gravitomagnetic moment and
  nonuniversality of the axial vortical effect at finite temperature}, Phys.
  Rev. D 103~(11) (2021) 116005.
\newblock \href {http://arxiv.org/abs/2102.01676} {\path{arXiv:2102.01676}},
  \href {https://doi.org/10.1103/PhysRevD.103.116005}
  {\path{doi:10.1103/PhysRevD.103.116005}}.

\bibitem{Landsteiner:2011cp}
K.~Landsteiner, E.~Megias, F.~Pena-Benitez, {Gravitational Anomaly and
  Transport}, Phys. Rev. Lett. 107 (2011) 021601.
\newblock \href {http://arxiv.org/abs/1103.5006} {\path{arXiv:1103.5006}},
  \href {https://doi.org/10.1103/PhysRevLett.107.021601}
  {\path{doi:10.1103/PhysRevLett.107.021601}}.

\bibitem{Jensen:2012kj}
K.~Jensen, R.~Loganayagam, A.~Yarom, {Thermodynamics, gravitational anomalies
  and cones}, JHEP 02~(2013) 088.
\newblock \href {http://arxiv.org/abs/1207.5824} {\path{arXiv:1207.5824}},
  \href {https://doi.org/10.1007/JHEP02(2013)088}
  {\path{doi:10.1007/JHEP02(2013)088}}.

\bibitem{Stone:2018zel}
M.~Stone, J.~Kim, {Mixed Anomalies: Chiral Vortical Effect and the Sommerfeld
  Expansion}, Phys. Rev. D 98~(2) (2018) 025012.
\newblock \href {http://arxiv.org/abs/1804.08668} {\path{arXiv:1804.08668}},
  \href {https://doi.org/10.1103/PhysRevD.98.025012}
  {\path{doi:10.1103/PhysRevD.98.025012}}.

\bibitem{Prokhorov:2022udo}
G.~Y. Prokhorov, O.~V. Teryaev, V.~I. Zakharov, {Hydrodynamic Manifestations of
  Gravitational Chiral Anomaly}, Phys. Rev. Lett. 129~(15) (2022) 151601.
\newblock \href {http://arxiv.org/abs/2207.04449} {\path{arXiv:2207.04449}},
  \href {https://doi.org/10.1103/PhysRevLett.129.151601}
  {\path{doi:10.1103/PhysRevLett.129.151601}}.

\bibitem{Becattini:2020sww}
F.~Becattini, {Polarization in relativistic fluids: a quantum field theoretical
  derivation}, Lect. Notes Phys. 987 (2021) 15--52.
\newblock \href {http://arxiv.org/abs/2004.04050} {\path{arXiv:2004.04050}},
  \href {https://doi.org/10.1007/978-3-030-71427-7_2}
  {\path{doi:10.1007/978-3-030-71427-7_2}}.

\bibitem{Vasak:1987um}
D.~Vasak, M.~Gyulassy, H.~T. Elze, {Quantum Transport Theory for Abelian
  Plasmas}, Annals Phys. 173 (1987) 462--492.
\newblock \href {https://doi.org/10.1016/0003-4916(87)90169-2}
  {\path{doi:10.1016/0003-4916(87)90169-2}}.

\bibitem{Hayata:2015lga}
T.~Hayata, Y.~Hidaka, T.~Noumi, M.~Hongo, {Relativistic hydrodynamics from
  quantum field theory on the basis of the generalized Gibbs ensemble method},
  Phys. Rev. D 92~(6) (2015) 065008.
\newblock \href {http://arxiv.org/abs/1503.04535} {\path{arXiv:1503.04535}},
  \href {https://doi.org/10.1103/PhysRevD.92.065008}
  {\path{doi:10.1103/PhysRevD.92.065008}}.

\bibitem{Stewart:2017zsu}
E.~Stewart, K.~Tuchin, {Magnetic field in expanding quark-gluon plasma}, Phys.
  Rev. C 97~(4) (2018) 044906.
\newblock \href {http://arxiv.org/abs/1710.08793} {\path{arXiv:1710.08793}},
  \href {https://doi.org/10.1103/PhysRevC.97.044906}
  {\path{doi:10.1103/PhysRevC.97.044906}}.

\bibitem{Huang:2011dc}
X.-G. Huang, A.~Sedrakian, D.~H. Rischke, {Kubo formulae for relativistic
  fluids in strong magnetic fields}, Annals Phys. 326 (2011) 3075--3094.
\newblock \href {http://arxiv.org/abs/1108.0602} {\path{arXiv:1108.0602}},
  \href {https://doi.org/10.1016/j.aop.2011.08.001}
  {\path{doi:10.1016/j.aop.2011.08.001}}.

\bibitem{Hongo:2020qpv}
M.~Hongo, K.~Hattori, {Revisiting relativistic magnetohydrodynamics from
  quantum electrodynamics}, JHEP 02~(2021) 011.
\newblock \href {http://arxiv.org/abs/2005.10239} {\path{arXiv:2005.10239}},
  \href {https://doi.org/10.1007/JHEP02(2021)011}
  {\path{doi:10.1007/JHEP02(2021)011}}.

\bibitem{Becattini:2012tc}
F.~Becattini, {Covariant statistical mechanics and the stress-energy tensor},
  Phys. Rev. Lett. 108 (2012) 244502.
\newblock \href {http://arxiv.org/abs/1201.5278} {\path{arXiv:1201.5278}},
  \href {https://doi.org/10.1103/PhysRevLett.108.244502}
  {\path{doi:10.1103/PhysRevLett.108.244502}}.

\bibitem{Becattini:2021suc}
F.~Becattini, M.~Buzzegoli, A.~Palermo, {Spin-thermal shear coupling in a
  relativistic fluid}, Phys. Lett. B 820 (2021) 136519.
\newblock \href {http://arxiv.org/abs/2103.10917} {\path{arXiv:2103.10917}},
  \href {https://doi.org/10.1016/j.physletb.2021.136519}
  {\path{doi:10.1016/j.physletb.2021.136519}}.

\bibitem{Becattini:2021iol}
F.~Becattini, M.~Buzzegoli, G.~Inghirami, I.~Karpenko, A.~Palermo, {Local
  Polarization and Isothermal Local Equilibrium in Relativistic Heavy Ion
  Collisions}, Phys. Rev. Lett. 127~(27) (2021) 272302.
\newblock \href {http://arxiv.org/abs/2103.14621} {\path{arXiv:2103.14621}},
  \href {https://doi.org/10.1103/PhysRevLett.127.272302}
  {\path{doi:10.1103/PhysRevLett.127.272302}}.

\bibitem{Liu:2021uhn}
S.~Y.~F. Liu, Y.~Yin, {Spin polarization induced by the hydrodynamic
  gradients}, JHEP 07~(2021) 188.
\newblock \href {http://arxiv.org/abs/2103.09200} {\path{arXiv:2103.09200}},
  \href {https://doi.org/10.1007/JHEP07(2021)188}
  {\path{doi:10.1007/JHEP07(2021)188}}.

\bibitem{Fu:2021pok}
B.~Fu, S.~Y.~F. Liu, L.~Pang, H.~Song, Y.~Yin, {Shear-Induced Spin Polarization
  in Heavy-Ion Collisions}, Phys. Rev. Lett. 127~(14) (2021) 142301.
\newblock \href {http://arxiv.org/abs/2103.10403} {\path{arXiv:2103.10403}},
  \href {https://doi.org/10.1103/PhysRevLett.127.142301}
  {\path{doi:10.1103/PhysRevLett.127.142301}}.

\bibitem{Liu:2020dxg}
S.~Y.~F. Liu, Y.~Yin, {Spin Hall effect in heavy-ion collisions}, Phys. Rev. D
  104~(5) (2021) 054043.
\newblock \href {http://arxiv.org/abs/2006.12421} {\path{arXiv:2006.12421}},
  \href {https://doi.org/10.1103/PhysRevD.104.054043}
  {\path{doi:10.1103/PhysRevD.104.054043}}.

\bibitem{Fu:2022myl}
B.~Fu, L.~Pang, H.~Song, Y.~Yin, {Signatures of the spin Hall effect in hot and
  dense QCD matter}, arXiv preprint (1 2022).
\newblock \href {http://arxiv.org/abs/2201.12970} {\path{arXiv:2201.12970}}.

\bibitem{Ivanov:2022geb}
Y.~B. Ivanov, A.~A. Soldatov, {On ambiguity of definition of shear and
  spin-hall contributions to $\Lambda$ polarization in heavy-ion collisions},
  Pisma Zh. Eksp. Teor. Fiz. 116~(3) (2022) 137--138.
\newblock \href {http://arxiv.org/abs/2206.06927} {\path{arXiv:2206.06927}},
  \href {https://doi.org/10.1134/S0021364022601300}
  {\path{doi:10.1134/S0021364022601300}}.

\bibitem{Palermo:2021hlf}
A.~Palermo, M.~Buzzegoli, F.~Becattini, {Exact equilibrium distributions in
  statistical quantum field theory with rotation and acceleration: Dirac
  field}, JHEP 10 (2021) 077.
\newblock \href {http://arxiv.org/abs/2106.08340} {\path{arXiv:2106.08340}},
  \href {https://doi.org/10.1007/JHEP10(2021)077}
  {\path{doi:10.1007/JHEP10(2021)077}}.

\bibitem{book:SokolovAndTernov}
I.~M.~T. A.~A.~Sokolov, Radiation from Relativistic Electrons (American
  Institute of Physics Translation Series), 2nd Edition, American Institute of
  Physics, 1986.

\bibitem{Hakim:1977}
R.~D. Tenreiro, R.~Hakim,
  \href{https://link.aps.org/doi/10.1103/PhysRevD.15.1435}{Transport properties
  of the relativistic degenerate electron gas in a strong magnetic field:
  Covariant relaxation-time model}, Phys. Rev. D 15 (1977) 1435--1447.
\newblock \href {https://doi.org/10.1103/PhysRevD.15.1435}
  {\path{doi:10.1103/PhysRevD.15.1435}}.
\newline\urlprefix\url{https://link.aps.org/doi/10.1103/PhysRevD.15.1435}

\bibitem{Chen:2012ca}
J.-W. Chen, S.~Pu, Q.~Wang, X.-N. Wang, {Berry Curvature and Four-Dimensional
  Monopoles in the Relativistic Chiral Kinetic Equation}, Phys. Rev. Lett.
  110~(26) (2013) 262301.
\newblock \href {http://arxiv.org/abs/1210.8312} {\path{arXiv:1210.8312}},
  \href {https://doi.org/10.1103/PhysRevLett.110.262301}
  {\path{doi:10.1103/PhysRevLett.110.262301}}.

\bibitem{Hidaka:2017auj}
Y.~Hidaka, S.~Pu, D.-L. Yang, {Nonlinear Responses of Chiral Fluids from
  Kinetic Theory}, Phys. Rev. D 97~(1) (2018) 016004.
\newblock \href {http://arxiv.org/abs/1710.00278} {\path{arXiv:1710.00278}},
  \href {https://doi.org/10.1103/PhysRevD.97.016004}
  {\path{doi:10.1103/PhysRevD.97.016004}}.

\bibitem{Dong:2020zci}
R.-D. Dong, R.-H. Fang, D.-F. Hou, D.~She, {Chiral magnetic effect for chiral
  fermion system}, Chin. Phys. C 44~(7) (2020) 074106.
\newblock \href {http://arxiv.org/abs/2001.05801} {\path{arXiv:2001.05801}},
  \href {https://doi.org/10.1088/1674-1137/44/7/074106}
  {\path{doi:10.1088/1674-1137/44/7/074106}}.

\bibitem{Fang:2016vpj}
R.-h. Fang, L.-g. Pang, Q.~Wang, X.-n. Wang, {Polarization of massive fermions
  in a vortical fluid}, Phys. Rev. C 94~(2) (2016) 024904.
\newblock \href {http://arxiv.org/abs/1604.04036} {\path{arXiv:1604.04036}},
  \href {https://doi.org/10.1103/PhysRevC.94.024904}
  {\path{doi:10.1103/PhysRevC.94.024904}}.

\bibitem{Gao:2019znl}
J.-H. Gao, Z.-T. Liang, {Relativistic Quantum Kinetic Theory for Massive
  Fermions and Spin Effects}, Phys. Rev. D 100~(5) (2019) 056021.
\newblock \href {http://arxiv.org/abs/1902.06510} {\path{arXiv:1902.06510}},
  \href {https://doi.org/10.1103/PhysRevD.100.056021}
  {\path{doi:10.1103/PhysRevD.100.056021}}.

\bibitem{Weickgenannt:2019dks}
N.~Weickgenannt, X.-L. Sheng, E.~Speranza, Q.~Wang, D.~H. Rischke, {Kinetic
  theory for massive spin-1/2 particles from the Wigner-function formalism},
  Phys. Rev. D 100~(5) (2019) 056018.
\newblock \href {http://arxiv.org/abs/1902.06513} {\path{arXiv:1902.06513}},
  \href {https://doi.org/10.1103/PhysRevD.100.056018}
  {\path{doi:10.1103/PhysRevD.100.056018}}.

\bibitem{Hattori:2019ahi}
K.~Hattori, Y.~Hidaka, D.-L. Yang, {Axial Kinetic Theory and Spin Transport for
  Fermions with Arbitrary Mass}, Phys. Rev. D 100~(9) (2019) 096011.
\newblock \href {http://arxiv.org/abs/1903.01653} {\path{arXiv:1903.01653}},
  \href {https://doi.org/10.1103/PhysRevD.100.096011}
  {\path{doi:10.1103/PhysRevD.100.096011}}.

\bibitem{Liu:2020ymh}
Y.-C. Liu, X.-G. Huang, {Anomalous chiral transports and spin polarization in
  heavy-ion collisions}, Nucl. Sci. Tech. 31~(6) (2020) 56.
\newblock \href {http://arxiv.org/abs/2003.12482} {\path{arXiv:2003.12482}},
  \href {https://doi.org/10.1007/s41365-020-00764-z}
  {\path{doi:10.1007/s41365-020-00764-z}}.

\bibitem{Dayi:2020uwx}
O.~F. Dayi, E.~Kilin\c{c}arslan, {Semiclassical transport equations of Dirac
  particles in rotating frames}, Phys. Rev. D 102~(4) (2020) 045015.
\newblock \href {http://arxiv.org/abs/2004.07510} {\path{arXiv:2004.07510}},
  \href {https://doi.org/10.1103/PhysRevD.102.045015}
  {\path{doi:10.1103/PhysRevD.102.045015}}.

\bibitem{Sheng:2020oqs}
X.-L. Sheng, Q.~Wang, X.-G. Huang, {Kinetic theory with spin: From massive to
  massless fermions}, Phys. Rev. D 102~(2) (2020) 025019.
\newblock \href {http://arxiv.org/abs/2005.00204} {\path{arXiv:2005.00204}},
  \href {https://doi.org/10.1103/PhysRevD.102.025019}
  {\path{doi:10.1103/PhysRevD.102.025019}}.

\bibitem{Gao:2020pfu}
J.-H. Gao, Z.-T. Liang, Q.~Wang, {Quantum kinetic theory for spin-1/2 fermions
  in Wigner function formalism}, Int. J. Mod. Phys. A 36~(01) (2021) 2130001.
\newblock \href {http://arxiv.org/abs/2011.02629} {\path{arXiv:2011.02629}},
  \href {https://doi.org/10.1142/S0217751X21300015}
  {\path{doi:10.1142/S0217751X21300015}}.

\bibitem{Landau:1969}
L.~D. Landau, E.~M. Lifshitz, Statistical physics part 1, 3rd Edition, Vol.~5,
  Pergamon Press, 1969.

\bibitem{Workman:2022ynf}
R.~L. Workman, Others, {Review of Particle Physics}, PTEP 2022 (2022) 083C01.
\newblock \href {https://doi.org/10.1093/ptep/ptac097}
  {\path{doi:10.1093/ptep/ptac097}}.

\bibitem{Deng:2012pc}
W.-T. Deng, X.-G. Huang, {Event-by-event generation of electromagnetic fields
  in heavy-ion collisions}, Phys. Rev. C 85 (2012) 044907.
\newblock \href {http://arxiv.org/abs/1201.5108} {\path{arXiv:1201.5108}},
  \href {https://doi.org/10.1103/PhysRevC.85.044907}
  {\path{doi:10.1103/PhysRevC.85.044907}}.

\bibitem{McLerran:2013hla}
L.~McLerran, V.~Skokov, {Comments About the Electromagnetic Field in Heavy-Ion
  Collisions}, Nucl. Phys. A 929 (2014) 184--190.
\newblock \href {http://arxiv.org/abs/1305.0774} {\path{arXiv:1305.0774}},
  \href {https://doi.org/10.1016/j.nuclphysa.2014.05.008}
  {\path{doi:10.1016/j.nuclphysa.2014.05.008}}.

\bibitem{Tuchin:2013apa}
K.~Tuchin, {Time and space dependence of the electromagnetic field in
  relativistic heavy-ion collisions}, Phys. Rev. C 88~(2) (2013) 024911.
\newblock \href {http://arxiv.org/abs/1305.5806} {\path{arXiv:1305.5806}},
  \href {https://doi.org/10.1103/PhysRevC.88.024911}
  {\path{doi:10.1103/PhysRevC.88.024911}}.

\bibitem{Inghirami:2016iru}
G.~Inghirami, L.~Del~Zanna, A.~Beraudo, M.~H. Moghaddam, F.~Becattini,
  M.~Bleicher, {Numerical magneto-hydrodynamics for relativistic nuclear
  collisions}, Eur. Phys. J. C 76~(12) (2016) 659.
\newblock \href {http://arxiv.org/abs/1609.03042} {\path{arXiv:1609.03042}},
  \href {https://doi.org/10.1140/epjc/s10052-016-4516-8}
  {\path{doi:10.1140/epjc/s10052-016-4516-8}}.

\bibitem{Vitiuk:2019rfv}
O.~Vitiuk, L.~V. Bravina, E.~E. Zabrodin, {Is different $\Lambda$ and $\bar
  \Lambda$ polarization caused by different spatio-temporal freeze-out
  picture?}, Phys. Lett. B 803 (2020) 135298.
\newblock \href {http://arxiv.org/abs/1910.06292} {\path{arXiv:1910.06292}},
  \href {https://doi.org/10.1016/j.physletb.2020.135298}
  {\path{doi:10.1016/j.physletb.2020.135298}}.

\bibitem{Ambrus:2020oiw}
V.~E. Ambrus, M.~N. Chernodub, {Hyperon\textendash{}anti-hyperon polarization
  asymmetry in relativistic heavy-ion collisions as an interplay between chiral
  and helical vortical effects}, Eur. Phys. J. C 82~(1) (2022) 61.
\newblock \href {http://arxiv.org/abs/2010.05831} {\path{arXiv:2010.05831}},
  \href {https://doi.org/10.1140/epjc/s10052-022-10002-y}
  {\path{doi:10.1140/epjc/s10052-022-10002-y}}.

\bibitem{Ambrus:2019ayb}
V.~E. Ambrus, {Helical massive fermions under rotation}, JHEP 08~(2020)
  016.
\newblock \href {http://arxiv.org/abs/1912.09977} {\path{arXiv:1912.09977}},
  \href {https://doi.org/10.1007/JHEP08(2020)016}
  {\path{doi:10.1007/JHEP08(2020)016}}.

\end{thebibliography}

\end{document}